\documentstyle[12pt, bezier]{article}
\newcommand{\lab}[1]{\label{#1}}

\newcommand{\rf}[1]{(\ref{#1})}
\newcommand{\fullref}[1]{\ref{#1} on page~\pageref{#1}}

\newcommand\beq{\begin{equation}}
\newcommand\eeq{\end{equation}}
\newcommand\beqy{\begin{eqnarray}}
\newcommand\eeqy{\end{eqnarray}}
\newcommand\beqys{\begin{eqnarray*}}
\newcommand\eeqys{\end{eqnarray*}}

\newtheorem{Conj}{Conjecture}
\newtheorem{Th}{Theorem}

\newtheorem{Stat}{Statement}

\newcommand{\reali}{{\hbox{{\rm I}\kern-.2em\hbox{\rm R}}}}
\newcommand{\complessi}{{\ \hbox{{\rm I}\kern-.6em\hbox{\bf C}}}}
\newcommand\de{\partial}

\newcommand{\sgn}{{\rm sgn}}
\newcommand{\lk}{{\rm lk}}
\newcommand{\slk}{{\rm slk}}
\newcommand{\xnot}{{x_0}}
\newcommand{\ave}[1]{\left\langle{#1}\right\rangle}

\newcommand{\dg}{{{\dag}}}
\newcommand\braket[2]{\left\langle{ {#1}\,,\,{#2} }\right\rangle}
\newcommand{\dphi}[1]{\frac{\overrightarrow\delta}{\delta\Phi^{#1}}}
\newcommand{\dphib}[1]{\frac{\overleftarrow\delta}{\delta\Phi^{#1}}}

\newcommand{\dphid}[1]{\frac{\overrightarrow\delta}{\delta
\Phi^\dg_{#1}}}
\newcommand{\dphidb}[1]{\frac{\overleftarrow\delta}{\delta
\Phi^\dg_{#1}}}

\newcommand{\antib}[2]{ \left({{#1}\,,\,{#2}}\right) }

\newcommand{\cl}{{\rm cl}}
\newcommand{\lpsi}{{ {\cal L}_\Psi }}

\newcommand{\domega}{d_\omega}
\newcommand{\domegab}{\overline d_\omega}
\newcommand{\calA}{{\cal A}}
\newcommand{\calB}{{\cal B}}

\newcommand{\calBtilde}{{\widetilde{\cal B}}}
\newcommand{\calO}{{\cal O}}
\newcommand{\calP}{{\cal P}}
\newcommand{\calS}{{\cal S}}
\newcommand{\sigmao}{\sigma^\omega}
\newcommand{\Omegao}{\Omega^\omega}
\newcommand{\varphib}{\overline\varphi}
\newcommand{\calT}{{\cal T}}
\newcommand{\calU}{{\cal U}}

\newcommand{\calM}{{\cal M}}
\newcommand{\gammatot}{{\gamma^{(\Sigma_K,K,\xnot)}}}
\newcommand{\gammatotp}[1]{{\gamma^{(\Sigma_K,K,\xnot)}_{#1}}}
\newcommand{\intcdue}{\int_{C_2(K\backslash\xnot)}}

\newcommand{\intsig}{\int_{\Sigma_K}}
\newcommand{\gf}{{\rm g.f.}}
\newcommand{\Ctilde}[2]{{\widetilde{C}_{#1}^{#2}}}
\newcommand{\sfram}{{\rm s.f.}}
\newcommand{\gammaSK}{{\gamma^{\Sigma_K}}}
\newcommand{\gammaKnot}{{\gamma^{(K,\xnot)}}}
\newcommand{\gammaSnot}{{\gamma^{(\Sigma_K,\xnot)}}}
\newcommand{\avel}[1]{{\ave{#1}_\lambda}}
\newcommand{\tildeU}{{\widetilde U}}
\newcommand{\Phior}{{\Phi'_\bot}}
\newcommand{\univec}[1]{{\frac{{#1}}{|{#1}|}}}

\newcommand\qq{\rm}
\newcommand\cmp[1]{{\qq Commun.\ Math.\ Phys.\ \bf #1}}
\newcommand\jmp[1]{{\qq J.\ Math.\ Phys.\ \bf #1}}
\newcommand\pl[1]{{\qq Phys.\ Lett.\ \bf #1}}
\newcommand\np[1]{{\qq Nucl.\ Phys.\ \bf #1}}
\newcommand\mpl[1]{{\qq Mod.\ Phys.\ Lett.\ \bf #1}}

\newcommand\lmp[1]{{\qq Lett.\ Math.\ Phys.\ \bf #1}}

\newcommand\cqg[1]{{\qq Class.\ Quant.\ Grav.\ \bf #1}}

\newcommand\tmp[1]{{\qq Theor.\ Math.\ Phys.\ \bf #1}}
\newcommand\anp[1]{{\qq Ann.\ Phys.\ \bf #1}}
\newcommand\anm[1]{{\qq Ann.\ Math.\ \bf #1}}
\newcommand\adm[1]{{\qq Adv.\ Math.\ \bf #1}}
\newcommand\rms[1]{{\qq Russ.\ Math.\ Surveys \bf #1}}
\newcommand\jp[1]{{\qq J.\ Phys.\ \bf #1}}
\newcommand\bAMS[1]{{\qq Bull.\ Amer.\ Math.\ Soc.\ \bf #1}}
\newcommand\tAMS[1]{{\qq Trans.\ Amer.\ Math.\ Soc.\ \bf #1}}

\begin{document}
\begin{titlepage}
\title{Abelian $BF$ Theories and Knot Invariants}
\author{
Alberto S. Cattaneo\\[10pt]
Lyman Laboratory of Physics\\
Harvard University\\
{\sc Cambridge, MA 02138, USA}\\
E-mail: {\tt cattaneo@math.harvard.edu}
}
\date{September 25, 1996\\[10pt]
{\sf HUTMP-96/B355, hep-th/9609205}}
%\\[10pt]
%P.A.C.S. 02.40, 11.15, 04.60}

\maketitle

\begin{abstract}
In the context of the Batalin--Vilkovisky formalism,
a new observable for the Abelian $BF$ theory is proposed
whose vacuum expectation value is related to the 
Alexander--Conway polynomial. The three-dimensional case is
analyzed explicitly, and it is proved to be anomaly free.
Moreover, at the second order in perturbation theory, a new 
formula for the second coefficient of the Alexander--Conway polynomial 
is obtained.
An account on the higher-dimensional generalizations is also given.
\end{abstract}
\end{titlepage}
\section{Introduction}
In recent years, the study of three-dimensional {\em topological
quantum field theories} (TQFT) has shed new light on knot invariants.

The non perturbative analysis of the Chern--Simons theory in \cite{WittCS}
has given an intrinsically three-dimensional definition of 
the Jones \cite{Jon} and HOMFLY \cite{HOMFLY}
polynomials, while the approach of \cite{FK} has shown a more direct
connection with the $(2+1)$-dimensional formulation.

Later, the perturbative expansion in the covariant gauge 
\cite{GMM,BN-th}
has shown that numerical knot invariants can be obtained in terms of integrals
over copies of the knot times copies of $\reali^3$, and the second-order
invariant has been computed explicitly.

A rigorous mathematical formulation of these integrals
has been given in \cite{BT} where
some subtleties (``anomalies'') that arise in this framework have also 
been pointed out.

Another three-dimensional TQFT is the so-called $BF$ theory \cite{Schw}.
The version with a cosmological term gives results that are equivalent
to those obtained in Chern--Simons theory 
\cite{C,CCFM}. The version
without cosmological term (or {\em pure}) admits an observable 
\cite{CCM,Cat,Lon}
whose {\em vacuum expectation value} (v.e.v.) is related to the
Alexander--Conway polynomial \cite{AC}.

This is a ``classical'' knot invariant that can be defined in any
odd dimension; yet it is quite difficult to find a good generalization
of the three-dimensional observable of the pure $BF$ theory in 
higher dimensions (s. \cite{CoM} for an attempt).

However, pure $BF$ theory is essentially an Abelian theory, as one can
see by rescaling $A\to\epsilon A$ and $B\to B/\epsilon$ since, in the
limit $\epsilon\to0$, the non-Abelian perturbation $B\wedge A\wedge A$ gets killed. 
Therefore, we expected to get the same invariants studying  the simpler
Abelian version of $BF$ theory. 

In this framework, we have got to define a new non-trivial observable
which, though rather involved, has a natural generalization in any odd
dimension. 

In the three-dimensional case, we show that this observable
is not ``anomalous" both in the usual field-theoretical 
and in the Bott and Taubes's \cite{BT} meaning; i.e., we prove
both that a quantum observable corresponding to the classical one
exists and that the topological nature of
its v.e.v.\ is not spoiled by the collapse of more than
two points together ({\em hidden faces}).

Moreover, we show that this v.e.v.\ yields, at the second order in 
perturbation theory, a new integral expression for the second 
coefficient of 
the Alexander--Conway polynomial (our conjecture is that the whole 
v.e.v.\ is the inverse of the Alexander--Conway polynomial).

More generally, our approach suggests a new way of defining knot
invariants (or even higher-degree forms on the space of imbeddings)
as integrals over copies of the knot and a cobounding surface that
turn out not to depend on the choice of the surface.

Eventually, we recall that the Abelian $BF$ theory has a physical 
application \cite{FGM}
as a tool for studying the bosonization of many-body
systems. It would be interesting to see if our observable has a physical
interpretation as well.

\subsection{Plan of the paper}
Explicitly, in the three-dimensional case, we want to consider the
classical action
\[
S_\cl = \int_{\reali^3} B\wedge dA,
\]
where $A$ and $B$ are one-forms. This theory is invariant under
adding an exact form to either $A$ or $B$ . A classical observable
for this theory is then given by
\[
\gammatotp\cl = \intsig B\wedge A 
+\frac12 \oint_{x<y\in K} [A(x)\,B(y)-B(x)\,A(y)],
\]
where $K=\de\Sigma_K$ is a knot.

This observable can be shown to be {\em on-shell} invariant
under the symmetries of the action, i.e., to be invariant up to terms
containing
the equations of motion (which state that $A$ and $B$ are closed forms).

Observables that are invariant only on shell can be considered in
the {\em Batalin--Vilkovisky} (BV) formalism \cite{BV}, which we briefly
review in Sec.\ \ref{sec-BV}.

The application of the BV formalism to the three-dimensional Abelian
$BF$ theory is given in Sec.\ \ref{tdabft}. (In App.\ \ref{app-zm},
we discuss the generalization to the case where there exist harmonic
zero modes.)

In Sec.\ \ref{sec-obs}, we discuss the BV formulation
of our observable and prove that its exponential is not anomalous,
i.e., it is possible to make it an observable by adding higher-order 
corrections in the Planck constant. (In App.\ \ref{app-corr}, 
we find the lowest-order correction explicitly.)

In Sec.\ \ref{sec-comp}, we describe the Feynman diagrams of the theory,
discuss the generalization to our case of the ``regularization" 
framework developed in \cite{BT}, and obtain the simplest---but 
non-trivial---numerical knot invariant that our theory produces, 
viz., the second coefficient of the Alexander--Conway polynomial.
(In App.\ \ref{app-hf}, we prove that this v.e.v.\ is not anomalous
in the sense of \cite{BT}, and generalize this result to any
order in perturbation theory.)

Finally, in Sec.\ \ref{sec-glim}, we describe the steps to be taken to 
define the higher-dimensional generalizations of our theory.

\section{The BV formalism}
\lab{sec-BV}
The BV formalism \cite{BV} is a generalization of the BRST formalism
\cite{BRST} which is applicable to theories whose symmetry closes
only on shell.  Moreover, even in the case of off-shell closed 
symmetries, the BV formalism allows dealing with observables that
are invariant only on shell, as the ones we are interested in.

In this section we give only a brief introduction.
We refer to Ref.\ \cite{Ans}, whose notations we follow,
for a thorough exposition of the BV formalism, as well as for a 
clear-cut discussion of the renormalization issue (which we have not
to deal with in the present paper since the theory we consider is
topological as well as Gaussian).

\subsection{Preliminaries}
We denote by $\Phi^i$ the {\em fields} one needs in a theory
(i.e., the physical fields, the ghosts, the antighosts, the Lagrange
multipliers and, if necessary, the ghosts for ghosts and so on);
the space of fields is called the {\em configuration space}.  We denote
by $\epsilon(\Phi^i)$, or simply by $\epsilon_i$, the ghost number
of the field $\Phi^i$.  By simplicity, we consider only the case
of a theory whose physical fields are bosonic, so the Grassmann
parity of $\Phi^i$ is given by $(-1)^{\epsilon_i}$.

In the BV formalism,
along with every field $\Phi^i$ one introduces an {\em antifield}
$\Phi^\dg_i$ with the same characteristics of its partner but the
ghost number, which instead is given by
\beq
\epsilon(\Phi_i^\dg) = -\epsilon_i - 1;
\lab{ghn}
\eeq
this also implies that the Grassmann parity is reversed.
The space of fields and antifields is called the {\em phase space}
\cite{SchwBV}.
In the next sections we will also use the antifields $\Phi^*_i$ 
satisfying
\beq
\Phi^\dg_i := *\Phi^*_i,
\lab{stardagger}
\eeq
where $*$ is the Hodge operator.

Over the phase space one introduces a supersymplectic 
structure \cite{SchwBV} which
allows the definition of the BV {\em antibracket}
\beq
\antib XY := X \braket{\dphib i}{\dphid i} Y
- X \braket{\dphidb i}{\dphi i} Y
\lab{defantib}
\eeq
and the BV {\em Laplacian}
\beq
\Delta X := \sum_i (-1)^{\epsilon_i+1} X \braket{\dphib i}{\dphidb i}.
\lab{deflap}
\eeq
Here $X$ and $Y$ are functionals over the phase space and
$\braket\cdot\cdot$ denotes the scalar product
\beq
\braket\alpha\beta := \int_M \alpha\wedge*\beta
\eeq
and $M$ is the manifold over which the theory is defined.

In Ref.\ \cite{WittBV}, the BV phase space is interpreted
as the tangent space $T\calM$ over the space of fields $\calM$.
In the finite-dimensional case, this is locally isomorphic to the
cotangent space $T^*\calM$ by using a volume form on $\calM$.
Then the Laplacian $\Delta$ on $T\calM$ is in correspondence with
the exterior derivative on $T^*\calM$.
However, the product of two functionals on $T\calM$ does not correspond
to the wedge product on $T^*\calM$, and the antibracket measures
this failure since
\beq
\antib XY = (-1)^{\epsilon(Y)} [\Delta(XY)-X\,\Delta Y -
(-1)^{\epsilon(Y)} \Delta X\ Y]
\eeq
(notice that the Laplacian as defined in \rf{deflap} acts from the 
right). 

Of course, in the infinite-dimensional case (in which we are 
interested), this description is only formal. In particular, the 
Laplacian depends on the regularization we use to define the functional
integral.

\subsection{BV cohomologies}
One can define some cohomologies on the phase space,
with respect to the gradation provided by the ghost number.
Each cohomology is defined by a coboundary operator, i.e., a nilpotent
operator of ghost number one.

The simplest coboundary operator is the Laplacian itself.
The second interesting coboundary operator is
\beq
\Omega X := \antib X\Sigma - i\hbar \Delta X,
\lab{defOmega}
\eeq
where $\Sigma$, the {\em quantum action}, is a bosonic functional 
that has to satisfy the {\em quantum master equation}
\beq
\antib\Sigma\Sigma -2i\hbar\Delta\Sigma = 0
\lab{qme}
\eeq
for $\Omega$ to be nilpotent.
Notice that \rf{qme} is equivalent to asking the Gibbs weight
$\exp(i\Sigma/\hbar)$ to be $\Delta$-closed.

The third coboundary operator we consider is
\beq
\sigma X := \antib XS
\lab{defsigma}
\eeq
where $S$, the {\em action}, is a bosonic functional that has
to satisfy the {\em master equation}
\beq
\antib SS =0
\lab{me}
\eeq
for $\sigma$ to be nilpotent.

A particular case, which we encounter in this paper, is provided by
a $\Delta$-closed action $S$. In this case
\beq
\Delta\sigma + \sigma\Delta = 0,
\lab{Deltasigma}
\eeq
so $\Delta$ and $\sigma$ define a double complex.
Moreover, by \rf{qme}, $S$ is also a quantum action for any $\hbar$
and, as such, it defines an $\Omega$-cohomology.

The restriction of the operator $\sigma$ to the configuration
space defines a new operator,
\beq
sX := (\sigma X)_{\big|_{\Phi^\dg=0}},
\lab{defs}
\eeq
which can be shown to be nilpotent {\em on shell}, i.e., modulo the
solutions of
\beq
S_\cl\,\frac{\overleftarrow\delta}{\delta\Phi^i} = 0,
\lab{onshell}
\eeq
where
\beq
S_\cl = S_{\big|_{\Phi^\dg=0}}
\lab{Scl}
\eeq
is the {\em classical action}.
Thus, $s$ defines a cohomology on the configuration space on shell.
If $s$ is nilpotent also off shell, one says that the symmetry
closes.  In this case the BRST approach is available and $s$ is
actually the BRST operator. Notice, however, that by \rf{defs} 
$s$ is defined to act from the right; the usual BRST operator $s_l$ is
the corresponding operator acting from the left, and one has
\beq
s_l X = (-1)^{\epsilon(X)} s X.
\lab{defBRST}
\eeq

\subsection{BV quantization}
The interest in the $\Omega$-cohomology relies on the fact that one can
formally show that the class of observables whose v.e.v.'s are
gauge-fixing independent is given by the $\Omega$-closed
bosonic functionals modulo $\Omega$-exact terms. More precisely,
one introduces the {\em partition function}
\beq
Z_\Psi := \int_\lpsi e^{\frac i\hbar\Sigma},
\lab{defZ}
\eeq
where the {\em Lagrangian submanifold} $\lpsi$ is defined by the 
equations
\beq
\Phi^\dg_i = \dphi i\,\Psi(\Phi),
\lab{deflpsi}
\eeq
and $\Psi$, the {\em gauge-fixing fermion},
is a functional on the configuration space that
has ghost number $-1$.

In Ref.\ \cite{WittBV}, prescription \rf{deflpsi} is shown to amount
to selecting the top form in the functional on $T^*\calM$ that,
under the isomorphism between $T\calM$ and $T^*\calM$, corresponds
to the Gibbs weight $\exp(i\Sigma/\hbar)$.

The v.e.v.\ of a functional $X$ over the phase space is then defined as
\beq
\ave X_\Psi = \frac1{Z_\Psi}
\int_\lpsi e^{\frac i\hbar\Sigma}\, X.
\lab{defvev}
\eeq

By using the formal properties of the functional integration, one has
then the following \cite{AD,SchwBV}
\begin{Stat}
If $\Sigma$ satisfies the quantum master equation
\rf{qme}, then
\begin{enumerate}
\item the partition function
$Z_\Psi$ and the expectation values of $\Omega$-closed functionals
do not change under infinitesimal variations of the 
gauge-fixing fermion $\Psi$, and
\item the expectation value of an $\Omega$-exact functional vanishes.
\end{enumerate}
\lab{statBV}
\end{Stat}
In the finite-dimensional case, Statement \ref{statBV} becomes a
rigorous theorem.

One can also show that the definitions \rf{defZ} and \rf{defvev}
correspond to the usual ones in the BRST formalism 
whenever applicable.

The $\sigma$-cohomology is useful since it is given by the 
$\Omega$-cohomology in the limit $\hbar\to0$ and is much easier
to study. The idea is to solve the quantum master equation and to
study the $\Omega$-cohomology by an expansion in powers of $\hbar$.
Notice, however, that from an action satisfying \rf{me}
is not always possible to obtain a quantum action satisfying \rf{qme}
that, in the limit $\hbar\to0$, yields the starting action; if this
does not happen, one calls the theory {\em anomalous}.
Moreover, even if the theory is not anomalous, a
$\sigma$-closed functional of  ghost number zero
does not always produce an $\Omega$-closed 
functional that, in the limit $\hbar\to0$, yields the starting one.
A sufficient condition for both to happen is that the 
one-ghost-number $\sigma$-cohomology be trivial.

As explained before, the $s$-cohomology is the restriction of
the $\sigma$-co\-hom\-ol\-ogy to the configuration space on shell.
Since it is easier to study than the $\sigma$-cohomology, one
can study the latter by an expansion in antifields. Under some mild
assumptions \cite{VBF}, one can prove that the extension from
a classical action $S_\cl$ to an action $S$ satisfying \rf{me} and
\rf{Scl} exists and is unique modulo {\em canonical transformations}
(i.e., transformations on the phase space that preserve the supersymplectic
structure).

\section{The three-dimensional Abelian $BF$ theory}
\lab{tdabft}
In this section we apply the BV formalism to the theory defined
by the classical action
\beq
S_\cl^\omega = \int_M B\wedge \domega A,
\lab{Sclomega}
\eeq
where
\begin{itemize}
\item $M$ is a three-manifold;
\item $A$ and $B$ are fields taking values in $\Omega^1(M)$;
\item $\domega = d + i\omega$, and
\item $\omega$ is an external $d$-closed source in $\Omega^1(M)$
(thus, $\domega^2=0$).
\end{itemize}
In the particular case $\omega=0$, we will simply write $S_\cl$ and
speak of the {\em pure} theory.
We can also split $S_\cl^\omega$ as
\beq
S_\cl^\omega = S_\cl -i \gamma_\cl^\omega,
\lab{Sclomegasplit}
\eeq
with
\beq
\gamma_\cl^\omega = -\int_M B\wedge \omega\wedge A=
\int_M \omega\wedge B\wedge A,
\lab{defgammaomegacl}
\eeq
and see $\gamma_\cl^\omega$ as a perturbation of the pure classical
action.

If $H^1(M,\domega)$ is trivial (for the general case s.\ App. \ref{app-zm}),
the symmetries of this theory are simply given by
\beq
\begin{array}{cc}
s^\omega A = \domega c, & s^\omega c =0,\\
s^\omega B = \domegab \psi & s^\omega\psi=0,
\end{array}
\lab{defso}
\eeq
where $\domegab = d - i\omega$, and $c$ and $\psi$ are the ghosts, 
which take values
in $\Omega^0(M)$ and have ghost number one.
Notice that $\gamma^\omega_\cl$
is on-shell invariant under the symmetry \rf{defso} of the pure
theory.

\subsection{The BV action}
The BV action corresponding to \rf{Sclomega} is given by
\beq
S^\omega = \int_M
B\wedge \domega A + A^*\wedge \domega c + B^*\wedge \domegab \psi +
\bar c^* h_c + \bar\psi^* h_\psi,
\lab{Somega}
\eeq
where $\bar c$ and $\bar \psi$ are
the antighosts, and
$h_c$ and $h_\psi$ are the Lagrange multipliers. The antighosts
and the Lagrange multipliers take values in $\Omega^0(M)$; the former
have ghost number minus one, the latter have ghost number zero.
The additional terms in the antighosts and Lagrange multipliers
are necessary to gauge fix the theory. 
Notice that we have used here the antifields
$*$ instead of the antifields $\dg$, s.\ \rf{stardagger}. 

If the Laplacian
$\domega^*\domega+\domega\domega^*$ has zero modes, additional
terms are required; for simplicity we suppose that there are no
zero modes, i.e., we suppose that the cohomology $H^*(M,\domega)$ is
trivial (s.\ App.\ \ref{app-zm} for the case when $H^1(M,\domega)$
is not trivial).

It is not difficult to see that $S^\omega$ satisfies the master
equation \rf{me} for any closed one-form $\omega$.
Notice moreover that $\Delta S=0$, so $S$ also satisfies 
the quantum master equation \rf{qme}.
Thus, we can quantize the theory with Gibbs weight 
$\exp(iS^\omega/\hbar)$ for any $\hbar$.  In the following we will
set $\hbar=1$.  Notice moreover that the action 
$S^\omega$ does not require the
choice of a metric on $M$, so we expect its partition function to
be a topological invariant of $M$.

The $\sigmao$ operator \rf{defsigma} acts on the fields and antifields
as follows
\beq
\begin{array}{cccc}
\sigmao\psi^*=-\domega B^*, & \sigmao B^*=-\domega A, &
\sigmao A = \domega c, & \sigmao c=0,\\
\sigmao c^*=-\domegab A^*, & \sigmao A^*=-\domegab B, &
\sigmao B = \domegab\psi, & \sigmao\psi =0,
\end{array}
\lab{sigmaAB}
\eeq
\beq
\begin{array}{cccc}
\sigmao h_c^*=-\bar c^*, & \sigmao \bar c^* =0, &
\sigmao\bar c=h_c, & \sigmao h_c=0,\\
\sigmao h_\psi^*=-\bar\psi^*, & \sigmao\bar\psi^*=0, &
\sigmao\bar\psi=h_\psi, & \sigmao h_\psi=0.
\end{array}
\lab{sigmaanti}
\eeq

It is very useful to consider the following linear combinations
\beq
\begin{array}{lcccccccc}
&& (3,-2) && (2,-1) && (1,0) && (0,1)\\
\calA &=& -\psi^* &+& B^* &+& A &+& c,\\
\calB &=& -c^* &+& A^* &+& B &+& \psi;
\end{array}
\lab{defcalAB}
\eeq
where by $(i,j)$ we denote an $i$-form of ghost number $j$.
Notice that $\calA$ and $\calB$ have an overall (i.e., form plus ghost)
degree equal to one. By \rf{defcalAB}, we can rewrite the
action as
\beq
S^\omega = \int_M
\calB\wedge\domega\calA +
\bar c^* h_c + \bar\psi^* h_\psi.
\lab{Somegacal}
\eeq
Moreover, we can rewrite \rf{sigmaAB} as
\beq
\begin{array}{lcr}
\sigmao_l\calA &=& \domega\calA,\\
\sigmao_l\calB &=& \domegab\calB,
\end{array}
\lab{sigmacalAB}
\eeq
where $\sigmao_l$ is the operator corresponding to $\sigmao$ but
acting from the left (as the exterior derivative); notice that
\beq
\sigmao_l X = (-1)^{\epsilon(X)} \sigmao X.
\eeq
Following \rf{Sclomegasplit}, we can split $S^\omega$ as
\beq
S^\omega = S -i \gamma^\omega,
\lab{Somegasplit}
\eeq
where
\beq
\gamma^\omega = -\int_M\calB\wedge\omega\wedge\calA=
\int_M\omega\wedge\calBtilde\wedge\calA,
\lab{defgammaomega}
\eeq
where the operator $\tilde{}$ acts by changing sign to odd-ghost-number
terms.
The splitting \rf{Somegasplit} is very convenient since not
only do both $S^\omega$ and $S$ satisfy the quantum
master equation \rf{qme}, but we also have
\beqy
\antib{\gamma^\omega}S &=& 0,\lab{goS}\\
\antib{\gamma^\omega}{\gamma^\omega}&=&0,\lab{gogo}\\
\Delta\gamma^\omega&=&0\lab{Deltago}
\eeqy
[notice that, because of \rf{sigmaAB+},
\rf{goS} does not hold if $H^1(M,d)$
contains nontrivial elements besides $\omega$].

We can also split the Gibbs weight $\exp(iS^\omega)$ into the
Gibbs weight $\exp(iS)$ times the observable
\beq
\Gamma[\omega] = \exp\gamma^\omega,
\lab{defOomega}
\eeq
which we can prove to be $\Omega$-closed as a consequence of
\rf{goS}, \rf{gogo} and \rf{Deltago}.
Notice that, if $\omega$ is not trivial, the action $S$ as to
be modified as in App.\ \ref{app-zm} [with $b_1=1$ and
$\varphi_1=\varphib_1=v\,\omega'/\braket{\omega'}{\omega'}$,
where $\omega'=\omega+d\alpha$ and $d^*\omega'=0$].

By using these notations, it is easy to prove that the theory really
depends only on the cohomology class of $\omega$. In fact, if we
substitute $\omega$ with $\omega+df$, the action $S^\omega$ gets an
extra contribution
\beq
S^\omega \longrightarrow S^\omega + \Omegao T^f
\eeq
with
\beq
T^f=\int_M f\calBtilde\wedge\calA.
\eeq
By noticing that, $\Delta T^f=0$, 
$\antib{\Omegao T^f}{\Omegao T^f}=0$ and 
$\antib{\Omegao T^f}{T^f}=0$, we can show that
\beq
\exp(i\Omegao T^f) = 1 + \Omegao U^f,
\eeq
with
\beq
U^f = \sum_{n=1}^{\infty}\frac{i^n}{n!}\,
(\Omegao T^f)^{n-1}\,T^f.
\eeq
Since, by Statement \ref{statBV}, the v.e.v.\ of an $\Omegao$-exact
functional vanishes, we conclude the (formal) proof that the partition
function of the theory depends only on the cohomology class of
$\omega$.

\subsection{The quantization}
To quantize the theory, we have to choose a gauge-fixing fermion.
A convenient choice is
\beq
\Psi = \braket{\domega\bar c}A + 
\braket{\domegab\bar \psi}B.
\lab{defPsi}
\eeq
Notice that to gauge fix the theory we need to choose a metric
on $M$, but, by Statement \ref{statBV}, the partition function will not
depend on it.

By \rf{defZ} and \rf{deflpsi}, we have
\beq
Z[M,\omega] = \int[DA\,DB\,Dc\,D\bar c\,D\psi\,D\bar\psi\, 
Dh_c\,Dh_\psi] \exp{(i S^\omega_{\rm g.f.})},
\lab{Zomega}
\eeq
where
\beq
S^\omega_{\rm g.f.} = \int_M B\wedge \domega A +
\braket{\domega\bar c}{\domega c} +
\braket{\domegab\bar\psi}{\domegab \psi} +
\braket{h_c}{\domega^* A} + \braket{h_\psi}{\domegab^*B}.
\lab{Somegagf}
\eeq
The partition function \rf{Zomega} can then be computed by
using the zeta-function regularization of the determinants and yields
\cite{Schw}
\beq
Z[M,\omega] = \calT(M,\domega),
\lab{ZomegaT}
\eeq
where $\calT$ is the Ray--Singer torsion, which is a topological
invariant of the manifold $M$ and depends only on the cohomology
class of the closed one-form $\omega$. This explicit result confirms
the previously discussed formal arguments.
Notice that any multiple of $\omega$ is still a closed one-form; thus,
we can consider $Z[M,\lambda\omega]$ as well.

Now we recall that
the Ray--Singer torsion is equal to the Reidemeister
torsion \cite{CM} and the Reidemeister torsion of the complement
of a knot is proportional to the inverse of the Alexander--Conway 
polynomial
of the knot itself \cite{MT}; thus, we can see the inverse of the
Alexander--Conway polynomial as the partition function 
\rf{ZomegaT} of an Abelian $BF$ theory.
More precisely, we have the following (cfr.\ \cite{BG} and
\cite{Roz})
\begin{Th}
If $M=\reali^3\backslash{\rm Tub}(K)$, 
where ${\rm Tub}(K)$ is a tubular neighborhood of the knot 
$K\in\reali^3$,
and $\omega\in H^1(M)$ 
is such that 
\beq
\oint_{K_1}\omega=1,
\lab{c1omega}
\eeq
with
$K_1$ a closed circle wrapping around $K$ only once,
then 
\beq
\frac{Z[M,\lambda\omega]}{Z[M]}
=
\frac1{i\lambda}\frac{z(\lambda)}{\Delta(K;z(\lambda))},
\quad z(\lambda) = 2i\sin(\lambda/2),
\lab{aveOomega}
\eeq
where $\Delta(K;z)$ is the Alexander--Conway polynomial
satisfying the skein relation
\[
\Delta(K_+;z) - \Delta(K_-;z) = z\,\Delta(K_0;z)
\]
and normalized to one on the unknot.
\lab{thmOomega}
\end{Th}

\section{Observables for the pure theory}
\lab{sec-obs}
From now on, we will consider only the pure theory defined
by the action $S$ in \rf{Somega} with $\omega=0$.
We will look for observables (i.e., $\Omega$-closed 
zero-ghost-number functionals modulo $\Omega$-exact terms)
that are metric independent. By Statement \ref{statBV}, their
v.e.v.'s will give topological invariants (up to framing)
since the action is metric independent as well.

Our survey is not exhaustive; i.e., there could exist other more
involved, metric-independent observables that could lead to other
topological invariants.

\subsection{Loop observables}
The simplest observables one can build are
\beq
\gamma_A^K = \oint_K A,\quad
\gamma_B^K  = \oint_K B,
\eeq
where $K$ is an exact one-cycle [if $K$ were only closed,
these functionals would not be closed under \rf{sigmaAB+}].
These observables are always $\Omega$-exact:
\beq
\gamma_A^K = -\Omega\beta^*_\Sigma,\quad
\gamma_B^K = -\Omega\alpha^*_\Sigma,
\lab{alphaO}
\eeq
where
\beq
\alpha^*_\Sigma = \int_\Sigma A^*,\quad
\beta^*_\Sigma = \int_\Sigma B^*,
\eeq
and $\Sigma$ is a surface cobounding $K$.  Any function of $\gamma_A$
or $\gamma_B$ separately will be $\Omega$-exact, too.  To get a 
nontrivial observable, we have to pair them; e.g., we can consider
the observable
\beq
\tau[K_1,K_2] = \gamma_A^{K_1}\,\gamma_B^{K_2}.
\lab{deftau}
\eeq
By \rf{alphaO}, we can show that
\beq
\tau[K_1,K_2] = -i\Delta(\gamma_A^{K_1}\,\alpha^*_{\Sigma_2})
-\Omega(\gamma_A^{K_1}\,\alpha^*_{\Sigma_2}).
\eeq
Since
\beq
\Delta(\gamma_A^{K_1}\,\alpha^*_{\Sigma_2}) = 
-\int_{\Sigma_2}\omega_{K_1} = -\#(K_1,\Sigma_2)=
-\lk(K_1,K_2)
\eeq
(where $\omega_{K_1}$ is the Poincar\'e dual of $K_1$, 
$\#$ denotes the intersection number and $\lk$ the linking number),
we have
\beq
\ave{\tau[K_1,K_2]} = i\,\lk(K_1,K_2).
\lab{avetau}
\eeq
An explicit computation of the l.h.s.\ with the gauge-fixing
\rf{defPsi} actually gives Gauss's formula.

\subsection{Surface observables}
As we have seen,
the loop observables are rather trivial. A more interesting observable
can be built if $\dim H^1(M)=\dim H_2(M,\de M)=1$; viz., define
\beq
\gamma^\Sigma =
\int_\Sigma\calBtilde\wedge\calA=
\int_\Sigma(B\wedge A + B^*\psi + cA^*),
\lab{defgammaSigma}
\eeq
with $\Sigma\in H_2(M,\de M)$.
This observable is essentially the same as in \rf{defgammaomega}
with $\Sigma$ the Poincar\'e dual of $\omega$, so we know
that its exponential
\beq
\Gamma[\Sigma,\lambda] = \exp(\lambda\gamma^\Sigma),
\lab{expGS}
\eeq
is an observable as well which, up to $\Omega$-exact terms, 
depends only on the homology class of $\Sigma$.
Moreover, the splitting \rf{Somegasplit}, shows us that
\beq
\ave{\Gamma[\Sigma,\lambda]}_M =
\ave{\exp(\lambda\gamma^\omega)}_M =
\frac{Z[M,\lambda\omega]}{Z[M]}.
\lab{avegammaomega}
\eeq
In particular, this holds when $M$ is as in the hypotheses of 
Thm.\ \ref{thmOomega}. Thus, from the r.h.s.\ of \rf{aveOomega} we
can read the v.e.v.\ of $\Gamma[\Sigma,\lambda]$.
Notice that condition \rf{c1omega}
on $\omega$ requires its 
Poincar\'e dual $\Sigma$ to satisfy $\#(\Sigma,K_1)=1$.
Since any surface $\Sigma_K$ {\em spanning}
the knot $K$ (i.e., any oriented surface $\Sigma_K$ imbedded in $\reali^3$
such that $K$ is identical with the boundary of $\Sigma_K$, and the
orientation on $\Sigma_K$ induces the given orientation on $K$)
satisfies this property, we have the following
\begin{Th}
If $M=\reali^3\backslash{\rm Tub}(K)$ and 
$\Sigma=\Sigma_K\bigcap M\in H_2(M,\de M)$, then
\beq
\ave{\Gamma[\Sigma,\lambda]}_M =
\frac1{i\lambda}\frac{z(\lambda)}{\Delta(K;z(\lambda))},
\quad z(\lambda) = 2i\sin(\lambda/2).
\lab{aveOSigma}
\eeq
\lab{thmOSigma}
\end{Th}
Notice that a spanning surface $\Sigma_K$ always exists; e.g., 
we can take the Seifert surface.

An expansion in powers of $\lambda$ of the l.h.s.\ of 
\rf{aveOSigma} would give a representation of
the coefficients of the inverse of
the Alexander--Conway polynomial as Feynman diagrams involving
only bivalent vertices on $\Sigma$.
However,
the problem of finding the propagators in a manifold like
the one described above is very difficult.  
In the next subsection,
we will see how to recast the problem as the computation of
a v.e.v.\ in $\reali^3$.

\subsection{Surface-plus-knot observables}
\lab{ssec-spko}
From now on we work in $\reali^3$ and consider a knot $K$ together
with a spanning surface $\Sigma_K$. Thm.\ \ref{thmOSigma} suggests
to consider the v.e.v.\ of the exponential of $\gamma^{\Sigma_K}$.
However, since $\Sigma_K$ has a boundary, $\gamma^{\Sigma_K}$
is not $\sigma$-closed anymore; actually,
\beq
\sigma\gamma^{\Sigma_K} = \oint_K(\psi A-Bc).
\lab{sigmagammaSigma}
\eeq
Therefore, we have to find another functional depending on $K$
(so that it vanishes when $\Sigma_K$ is closed) such that
its $\sigma$-variation cancels \rf{sigmagammaSigma}. We first consider
\beq
\gamma^{(K,\xnot)} = \frac12\int_{x<y\in K} 
[A(x)\, B(y)-B(x)\, A(y)],
\lab{defgammaK}
\eeq
where $\xnot$ is a base point on $K$.
Notice that $K\backslash\xnot$ is diffeomorphic to $\reali$,
so its configuration spaces $C_n(K\backslash\xnot)$ are
diffeomorphic to the configuration spaces $C_n(\reali)$
described in \cite{FM} (s.\ also subsection \fullref{ssec-reg}).  
On these spaces it is possible to
introduce the {\em tautological forms}
\beq
\eta_{ij} = \phi^*_{ij}\,\omega, \quad i,j=1,\ldots,n,\quad
i\not= j,
\eeq
where $\phi^*_{ij}$ denotes the pullback via the map
\beq
\phi_{ij}(\vec x) = \sgn(x_i-x_j),
\quad \vec x\in C_n(\reali),
\eeq
and $\omega=1/2$ is the the unit volume element
on $S^0$.
With these notations, we can rewrite \rf{defgammaK} as
\beq
\gamma^{(K,\xnot)} =
\int_{C_2(K\backslash\xnot)}A_1\wedge\eta_{12}\wedge B_2.
\lab{AetaB}
\eeq
Now a simple computation shows that
\beq
\sigma\gamma^{(K,\xnot)} = -\oint_K(\psi A-Bc)+
\psi(\xnot)\,\oint_KA -\oint_KB\,c(\xnot);
\lab{sigmagammaK}
\eeq
so the first term cancels \rf{sigmagammaK}. We have then to find
another functional (vanishing when $\Sigma_K$ has no boundary)
whose variation cancels the second and third
terms. It is not difficult to see that
\beq
\gamma^{(\Sigma_K,\xnot)}=\psi(\xnot)\,\int_{\Sigma_K}B^*+
\int_{\Sigma_K}A^*\ c(\xnot)
\lab{defgamma0}
\eeq
does the job. Thus, we can define the following $\sigma$-closed
(actually, $\Omega$-closed) functional
\beq
\gammatot = \gamma^{\Sigma_K} + \gamma^{(K,\xnot)} +
\gamma^{(\Sigma_K,\xnot)}.
\lab{gammatot}
\eeq
In the case of links---which we will not consider anymore in the 
following---the observable has to be modified as
\beq
\gamma^{(\Sigma_K,K,\{\Sigma_{K_i},x_{0i}\})} =
\gamma^{\Sigma_K} + \sum_i\left[{
\gamma^{(K_i,x_{0i})} + \gamma^{(\Sigma_{K_i},x_{0i})}
}\right],
\eeq
where $\Sigma_K$ is a spanning surface for the link $K$ while
each $\Sigma_{K_i}$ is a spanning surface only for the component
$K_i$, whose base point is denoted by $x_{0i}$.

Then, recalling \rf{expGS},
we want to consider the exponential of $\gammatot$,
\beq
\calO_0[K,\lambda]=\exp(\lambda\gammatot),
\lab{calO0}
\eeq
which is $\sigma$-closed and hence a candidate to be an observable.
Actually,
\beq
\Delta\calO_0[K,\lambda]=\frac{\lambda^2}2\,\calO_0[K,\lambda]\,
\antib\gammatot\gammatot
\lab{DeltaO0}
\eeq
vanishes if we are working in {\em standard framing}, i.e., if
\beq
\slk(K)=\int_{\Sigma_K}\omega_K=0,
\lab{defsf}
\eeq
where $\omega_K$ is the Poincar\'e dual of $K$ and $\slk$ denotes
the self-linking number [whose definition via \rf{defsf} relies
on a choice of regularization].
With this hypothesis, we expect
the v.e.v.\ of $\calO_0$ not to depend on the gauge fixing and, as
a consequence, to be metric independent.

By essentially the same proof that led to the invariance (modulo
$\Omega$-exact terms) of $\Gamma[\omega]$, s.\ \rf{defOomega},
under $\omega\rightarrow\omega+d\eta$, 
we can prove that $\calO_0$ is invariant 
(modulo $\Omega$-exact terms) under $\Sigma_K\rightarrow\Sigma_K+
\de T$ with $T\in\Omega_3(\reali^3)$. 

From \rf{aveOSigma} we expect the v.e.v.\ of $\calO_0[K,\lambda]$ to be
proportional to the inverse of the Alexander--Conway polynomial.
The proportionality constant, which depends on $\lambda$, could be 
spoiled when we send $\reali^3\backslash{\rm Tub}(K)$ to
$\reali^3$; thus, we can make only the weaker statement that
\beq
\frac{\ave{\calO_0[K_\sfram,\lambda]}}
{\ave{\calO_0[\bigcirc_\sfram,\lambda]}}
= 
\frac1{\Delta(K;z(\lambda))},
\quad z(\lambda) = 2i\sin(\lambda/2),
\lab{aveOK0}
\eeq
where $K_\sfram$ and $\bigcirc_\sfram$ are, respectively, 
a generic knot and the unknot in standard framing.

This result has to be compared with the similar formulae
obtained in the context of the {\em non-Abelian} pure $BF$ theory 
\cite{CCM,Cat}. It should not amaze that the Abelian and
non-Abelian pure $BF$ theories are under this respect equivalent,
for, as observed in \cite{Cat}, the v.e.v.'s of the latter
can be computed exactly in saddle-point approximation (s.\ also
the observation in the Introduction).

We want to point out the precise meaning of \rf{aveOK0}:
We are not claiming that the coefficients of the $\lambda$-expansion of
$\ave{\calO_0}$ are a sum of numerical knot invariants up to factors
containing the self-linking number. We are saying that these numerical
knot invariants are well defined only if the knot is in
standard framing;
otherwise $\calO_0$ is not an observable, and
we are not guaranteed that its v.e.v.\ is a topological invariant.

This means that, to compute this v.e.v., we have to choose a particular
presentation of the knot, viz., one in which the
self-linking number vanishes, and that this v.e.v.\ should be invariant
only under deformations that do not change the self-linking number.
In the next section, we will discuss how to drop this cumbersome
condition.

\subsection{The corrected surface-plus-knot observable}
\lab{ssec-cspko}
For the purposes of this subsection, it is convenient to rescale
\beq
\calB\longrightarrow \calB/\lambda,
\lab{rescale}
\eeq
so the Gibbs weight becomes $\exp(iS/\lambda)$ and we recognize
$\lambda$ as the Planck constant of the theory.  Under this rescaling
we also have
\beq
\calO_0[K,\lambda]\longrightarrow \calO_0[K]=\exp\gammatot;
\eeq
thus, $\calO_0$ is a classical observable, i.e.,
it is $\sigma$-closed and does not depend on
the Planck constant.
Now we look for a quantum generalization
\beq
\calO[K,\lambda] = \sum_{n=0}^\infty (i\lambda)^n\,\calO_n[K]
\lab{defOK}
\eeq
satisfying
\beq
\Omega\calO=(\sigma-i\lambda\Delta)\calO=0,
\lab{OmegaO}
\eeq
and hence
\beq
\sigma\calO_{n} = \Delta\calO_{n-1},\quad\mbox{$n=1,2,\ldots$}.
\lab{Onn}
\eeq
Notice that a solution $O_{n}$, if it exists,
is defined only up to $\sigma$-closed
terms. However, if we send $O_n$ into $O_n+\sigma Y_n$, the 
solution $O_{n+1}$ will be sent into $O_{n+1}-\Delta Y_n$ because 
of \rf{Deltasigma}. Thus, $\calO$ will be changed by an $\Omega$-exact
term. On the other hand, if we add a nontrivial $\sigma$-closed
term to $O_n$, then this---together with the extra contribution 
$O_{n+1}$ receives by \rf{Onn}---lets $\calO$ get an extra 
$\Omega$-closed term. 
This means that the solution of \rf{OmegaO}, if it exists, is unique
only up to $\Omega$-closed terms, i.e., up to other observables.

It is not difficult to see, by \rf{Deltasigma}, 
that, if \rf{Onn} holds up to a fixed $n-1$, then
\beq
\sigma\Delta\calO_{n-1}=0.
\eeq
Thus, the r.h.s.\ of \rf{Onn} is $\sigma$-closed; however, to solve
\rf{Onn}, we want it to be $\sigma$-exact, which is not guaranteed.
If this happens, then the observable $\calO$ exists and we say
that $\calO_0$ is not anomalous. 

Among the possible solutions of \rf{OmegaO}, we are interested
in the ones that depend only on the triple 
$\Sigma_K,K,\xnot$ and that reduce to $\exp\gamma^{\Sigma_K}$
when $\Sigma_K$ has no boundary. We call these solutions {\em proper}.

Now notice that the action is invariant under the change of
variables $(\calA,\calB)\to(\calB,\calA)$ while the observable
$\gammatot$ is odd under it. This means that the corrections can
be chosen to have a well-defined parity. By induction, one can see
that they can be written as integrals of products of $\calB\wedge\calA$
and  $\calBtilde\wedge\calA$ over submanifolds of
products of configuration spaces of $\reali^3$.
Moreover, $\Delta\calO_n$ will have the same structure.
Since $\calBtilde\wedge\calA$ and $\calB\wedge\calA$
are overall two-forms (i.e., each component has form degree plus
ghost number equal to two), a product of them will be an overall
even form. As a consequence, the zero-ghost-number part will be an
even form, while the one-ghost-number part will be an odd form.
However, only the even homology spaces of
the configuration spaces of $\reali^3$ are nontrivial.
Therefore, since $\calO_n$ has ghost number zero, it can be
a non-trivial element of the $\sigma$-cohomology, whereas
$\Delta\calO_n$, which has ghost number one and is $\sigma$-closed, 
must be $\sigma$-exact.
Thus, we have proved the following
\begin{Th}
The classical observable $\calO_0[K]$ is not anomalous and
admits a proper extension.
\lab{th-notan}
\end{Th}

By Statement \ref{statBV}, we expect 
the v.e.v.\ of $\calO$ to be a topological invariant (up to framing) of 
the triple $\Sigma_K,K,\xnot$.
If the argument proving the invariance (modulo $\Omega$-exact
terms) under a deformation
$\Sigma_K\rightarrow\Sigma_K+\de T$ with $T\in\Omega_3(\reali^3)$
goes through, we arrive at the following
\begin{Conj}
The v.e.v.\ of a proper solution $\calO[K,\lambda]$ is a
regular-isotopy invariant of the knot $K$. 
\lab{conj-Kinv}
\end{Conj}

Now write this invariant as the sum of an ambient-isotopy 
invariant and a regular-isotopy invariant that vanishes in standard framing.
If the second contribution can be written as the v.e.v.\ of an $\Omega$-closed
observable, then we can redefine the proper solution $\calO$ by subtracting
this term; so we arrive to the following
\begin{Conj}
There exists a proper solution whose v.e.v.\ is an ambient-isotopy invariant.
\lab{conj-amb}
\end{Conj}

Eventually, since $\calO$ is a quantum generalization of $\calO_0$
to which it reduces in standard framing, and the v.e.v.\ of $\calO_0$
is expected to satisfy \rf{aveOK0}, we have our last
\begin{Conj}
The proper solution $\calO[K,\lambda]$ of Conjecture \ref{conj-amb}
satisfies
\beq
\frac{\ave{\calO[K,\lambda]}_\lambda}
{\ave{\calO[\bigcirc,\lambda]}_\lambda}= 
\frac1{\Delta(K;z(\lambda))},
\quad z(\lambda) = 2i\sin(\lambda/2),
\lab{aveOK}
\eeq
where $\bigcirc$ denotes the unknot.
\lab{conj-KAC}
\end{Conj}
If only Conjecture \ref{conj-Kinv} holds, we still expect \rf{aveOK} to hold
but only if the standard framing is chosen.

\subsubsection{The first correction}
In App.\ \ref{app-corr}, we discuss how to find the first
correction to $\calO_0$. In particular, we show that a 
proper solution is given by
\beq
\calO_1 = \calO_0\ U_1,
\lab{O1}
\eeq
with
\beq
U_1 = \Delta u_1,
\lab{Uu}
\eeq
and
\beq
u_1 = \gamma_{ABB}\,\int_{\Sigma_K} B^* +
      \gamma_{BAA}\,\int_{\Sigma_K} A^*,
\lab{u1}
\eeq
where
\beqy
\gamma_{ABB} &=& 
\int_{C_3(K\backslash\xnot)}A_1\eta_{12} B_2
\eta_{23}B_3
=-\frac12\int_{x<y<z\in K} B(x)\, A(y)\, B(z),
\lab{gammaABB}\\
\gamma_{BAA} &=& 
\int_{C_3(K\backslash\xnot)}B_1\eta_{12} A_2
\eta_{23}A_3
=-\frac12\int_{x<y<z\in K} A(x)\, B(y)\, A(z).
\lab{gammaBAA}
\eeqy
Remember that $\calO_1$ is defined up to a $\sigma$-closed term.
Our choice---\rf{O1}, \rf{Uu} and \rf{u1}---is 
particularly convenient since it gives the correct
v.e.v.\ of $\calO$ to the second order in $\lambda$.  
To see this, we first observe that any Wick contraction
in the computation of a v.e.v.\ carries a factor
$\lambda$. Thus, at order $\lambda^2$, the v.e.v.\ of $\calO_0+i\lambda
\calO_1$ will contain the v.e.v.'s
of $1/2(\gammatot)^2$ and of $i\lambda U_1$. 
Since $\Delta U_1=0$, we have
\beq
\Omega O_2=0,
\eeq
where
\beq
O_2=\left({
\frac12(\gammatot)^2+i\lambda U_1 
}\right);
\lab{O2}
\eeq
therefore, no other correction is needed to make this
second-order term an observable.

Notice that any redefinition of $U_1$ obtained by adding a 
$\Delta$-closed term will have the same property. By \rf{Onn},
this additional term must also be $\sigma$-closed, so it will
be $\Omega$-closed as well.  Thus, as expected, $O_2$ is defined
up to $\Omega$-closed terms whose v.e.v.'s are of order $\lambda^2$.
There exist only a few of such terms, i.e., $\lambda^2$, $\lambda\tau[K]$
and $\tau[K]^2$, where 
[cfr.\ \rf{deftau}]
\beq
\tau[K] = \gamma_A^K\,\gamma_B^K.
\lab{deftauK}
\eeq
By \rf{avetau}---and remembering \rf{rescale}---we see that
\beq
\ave{O_2+k\tau^2+i\lambda l\tau - \lambda^2 m}_\lambda=
\ave{O_2}_\lambda - \lambda^2
[2k(\slk K)^2+l\slk K + m];
\lab{aveO2tau}
\eeq
i.e., the second order of $\ave{\calO[K,\lambda]}_\lambda$ 
is defined up to a quadratic function of the self-linking number 
of $K$. We will see in the next section that, 
choosing $k=-3/16$ and $l=m=0$, 
Conjectures \ref{conj-Kinv}, \ref{conj-amb} and 
\ref{conj-KAC} hold at this order.

\paragraph{Remark} Notice that we are allowed to add to $O_2$ only
contributions of the form $\lambda$ times an observable, for
$\lambda$-independent contributions 
would change the classical part of the observable.
Thus, $\tau^2$ would not be an allowed contribution. However,
as shown at the end of App.\ \ref{app-corr}, adding $\tau^2$
to $O_2$ is equivalent to adding a $\lambda$-independent correction to
$U_1$, s.\ \rf{deltaU1}.

\section{Computation of v.e.v.'s}
\lab{sec-comp}
In this section we will describe the perturbative expansion of 
$\ave{\calO_0}_\lambda$ with the gauge fixing \rf{defPsi}.
We will also explicitly compute the second order term of $\ave{\calO_0}$
and $\ave{\calO}$. We will see that the latter satisfies, at this order,
Conjectures \ref{conj-Kinv} and \ref{conj-KAC}.

\subsection{Gauge fixing and propagators}
We will work in the covariant gauge fixing; i.e., we will choose the 
gauge-fixing fermion as in \rf{defPsi} with $\omega=0$. By \rf{deflpsi}
and \rf{stardagger}, this amounts to setting
\beq
\begin{array}{lr}
A^*=*d\bar c, & \bar c^*=*d^*A,\\
B^*=*d\bar\psi, & \bar\psi^*=*d^*B,
\lab{setstar}
\end{array}
\eeq
while all the other antifields are set to zero. Thus, the 
gauge-fixed action reads [cfr. \rf{Somegagf}]
\beq
S_{\rm g.f.} = \int_{\reali^3} B\wedge d A +
\braket{d\bar c}{d c} +
\braket{d\bar\psi}{d\psi} +
\braket{h_c}{d^* A} + \braket{h_\psi}{d^*B},
\lab{Sgf}
\eeq
from which we can read the propagators (we write only the ones we are
interested in)
\beq
\begin{array}{lcccl}
\ave{A_\mu(x)\,B_\nu(y)}_\lambda &=& 
\ave{B_\mu(x)\,A_\nu(y)}_\lambda &=&
i\lambda\,\frac1{4\pi}\,\epsilon_{\mu\nu\rho}\,
\frac{(x-y)^\rho}{|x-y|^3},\\
\ave{c(x)\,\bar c(y)}_\lambda &=&
-\ave{\bar c(x)\,c(y)}_\lambda &=&
i\lambda\,\frac1{4\pi}\,\frac1{|x-y|},\\
\ave{\psi(x)\,\bar\psi(y)}_\lambda &=&
-\ave{\bar\psi(x)\,\psi(y)}_\lambda &=&
i\lambda\,\frac1{4\pi}\,\frac1{|x-y|}.\\
\end{array}
\lab{props}
\eeq
Notice that the propagators are exactly the same as in Chern--Simons theory \cite{GMM,BN-th}. 

\subsubsection{Parity}
We have already observed that, under $(\calA,\calB)\to(\calB,\calA)$,
the action is left unchanged while $\gammatot$ changes sign.
As a consequence, in the gauge-fixing defined by \rf{setstar},
the propagators \rf{props} are invariant under 
\beq
(A,B,c,\bar c,\psi,\bar\psi,h_c,h_\psi)\rightarrow
(B,A,\psi,\bar\psi,c,\bar c,h_\psi,h_c).
\lab{parity}
\eeq
Thus, all the terms in perturbation theory that are odd under
\rf{parity} [like, e.g., the v.e.v.\  of $(\gammatot)^{2n+1}$]
will vanish.

\subsubsection{Supersymmetry}
Another observation that will simplify the discussion 
of the perturbation
theory concerns the supersymmetry of the action \rf{Sgf}
(which is the same that holds in its non-Abelian generalization 
\cite{MS}); viz.,  
we can define a fermionic vector operator $Q$, i.e.,
an operator satisfying
\beq
[Q_\alpha,Q_\beta]_+ = Q_\alpha\,Q_\beta+Q_\beta\,Q_\alpha=0,
\eeq
which annihilates the action:
\beq
Q S = 0.
\eeq
Actually, there exist two such operators (which, moreover, anticommute
with each other); the first one acts as
\beq
\begin{array}{lcl}
{}(QA)_{\alpha\beta} &=&
\epsilon_{\alpha\beta\gamma}\de^\gamma\bar\psi,\\
{}(QB)_{\alpha\beta} &=& \epsilon_{\alpha\beta\gamma}\de^\gamma\bar c,\\
{}(Qc)_\alpha &=& -A_\alpha,\\
{}(Q\psi)_\alpha &=& -B_\alpha,\\
{}(Q\bar c)_\alpha &=& 0,\\
{}(Q\bar\psi)_\alpha &=& 0,\\
{}(Qh_c)_\alpha &=& \de_\alpha\bar c,\\
{}(Qh_\psi)_\alpha &=& \de_\alpha\bar\psi.
\end{array}
\lab{defQ}
\eeq
The second operator is obtained by exchanging $(c,\bar c,h_c)$ with
$(\psi,\bar\psi,h_\psi)$.

A consequence of this supersymmetry is that the v.e.v.\ of a 
$Q$-exact functional vanishes. We want to point out that this 
supersymmetry is peculiar of $\reali^3$, but holds (with a different
$Q$-operator) also for other gauge fixings.

\subsubsection{Regularization}
\lab{ssec-reg}
The propagators \rf{props} diverge as the two points where the fields
are evaluated approach to each other. The non-regularized v.e.v.'s of
our observables are integrals of these propagators over products of
$C_k(K\backslash\xnot)$ and $\Sigma_K$. To avoid divergences we have
to give a prescription to split the points in these integrals.

Our choice will essentially follow the approach of \cite{BT} with some
important modifications due to the presence of the surface $\Sigma_K$
(s.\ also \cite{Lon}). The idea is to start defining the
Fulton--MacPherson \cite{FM}
compactification $C_n(\reali^3)$ of the configuration space 
of $n$ points in
$\overline{\reali^3}$, where the latter is the compactification of
$\reali^3$ obtained by replacing the infinity with its blow up.
Then, denoting by $B^2$ a two-dimensional surface whose boundary is
diffeomorphic to $S^1$,
we can consider the following imbeddings of compact spaces
\beq
{\rm pt}\hookrightarrow S^1 \hookrightarrow 
\overline{B^2} \hookrightarrow\overline{\reali^3},
\eeq
where $\rm pt$ is a base point on the sphere $S^1$ which is
mapped to the boundary of $B^2$.
This allows us to define the configuration space $C_n^t$ of $n$
points on the knot distinct from the base point and $t$ points on
its spanning surface. Notice that the points on the
knot can be ordered, so $C_n^t$ has $n!$ connected components.
We will denote by $\Ctilde nt$ the identity component (i.e., the 
component with points on $S^1$ ordered as $0,1,2,\ldots,n$).

Our regularization prescription to compute the v.e.v.'s will be
to replace $C_k(K\backslash\xnot)^n\times(\Sigma_K)^t$ with
$C_{kn}^t$.
Moreover, we will rewrite the propagators \rf{props} as
\beq
\begin{array}{lcccl}
\ave{A_i\wedge B_j}_\lambda &=& \ave{B_i\wedge A_j}_\lambda &=& 
i\lambda\,\theta_{ij},\\
\ave{c_i\,(*d\bar c)_j}_\lambda &=& 
\ave{(*d\bar c)_i\,c_j}_\lambda &=&
-i\lambda\,\theta_{ij},\\
\ave{\psi_i\,(*d\bar\psi)_j}_\lambda &=& 
\ave{(*d\bar \psi)_i\,\psi_j}_\lambda &=&
-i\lambda\,\theta_{ij},
\end{array}
\lab{propstheta}
\eeq
where $\theta_{ij}$ is the tautological form on $\reali^3$ defined
as the pullback of the $SO(3)$-invariant unit volume element on
$S^2$ by the map
\beq
\phi_{ij}(\vec x) = \frac{x_j-x_i}{|x_j-x_i|},
\quad\vec x\in C_n(\reali^3).
\eeq
Notice that $\theta_{ij}$ satisfies
\beq
d\theta_{ij}=0,\quad
\theta_{ij}^2 = 0,\quad \theta_{ji} = -\theta_{ij}.
\eeq

\begin{figure}
\unitlength 1.00mm
\linethickness{0.4pt}
\begin{picture}(140.00,55.00)
\put(0.00,5.00){\line(1,0){40.00}}
\put(50.00,5.00){\line(1,0){40.00}}
\put(100.00,5.00){\line(1,0){40.00}}
%\bezvec{224}(15.00,5.00)(25.00,31.00)(35.00,5.00)
\put(35.00,5.00){\vector(1,-3){0.2}}
\bezier{224}(15.00,5.00)(25.00,31.00)(35.00,5.00)
%\end
\put(5.00,5.00){\vector(1,2){4.50}}
\put(9.50,14.00){\vector(3,2){10.50}}
\put(20.00,21.00){\vector(1,-3){5.33}}
%\bezvec{408}(55.00,5.00)(65.00,55.00)(75.00,5.00)
\put(75.00,5.00){\vector(1,-4){0.2}}
\bezier{408}(55.00,5.00)(65.00,55.00)(75.00,5.00)
%\end
%\bezvec{244}(60.00,5.00)(65.00,35.00)(70.00,5.00)
\put(70.00,5.00){\vector(1,-4){0.2}}
\bezier{244}(60.00,5.00)(65.00,35.00)(70.00,5.00)
%\end
\put(65.00,5.00){\vector(1,2){10.00}}
\put(75.00,25.00){\vector(1,-2){10.00}}
\put(110.00,5.00){\vector(1,3){5.00}}
\put(115.00,20.00){\vector(1,-3){5.00}}
\put(125.00,5.00){\vector(-1,2){6.00}}
\put(119.00,17.00){\vector(4,-3){16.00}}
\put(100.00,5.00){\vector(3,1){15.00}}
\put(100.00,5.00){\vector(1,3){3.67}}
\put(103.67,16.00){\vector(4,-1){16.33}}
\put(120.00,11.92){\vector(3,4){7.56}}
\put(50.00,5.00){\vector(1,4){6.50}}
\put(56.50,31.00){\vector(3,-2){29.50}}
\end{picture}
\caption{Some examples of diagrams with nonvanishing
v.e.v.}\label{fig-examples}
\end{figure}
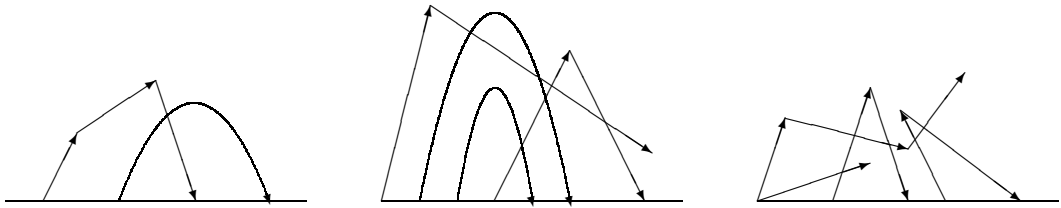

To compute our v.e.v.'s, we will have to integrate
these two-forms as well as the
tautological zero-forms $\eta_{ij}$ [appearing in \rf{AetaB}]
over some $C_n^t$.  If we choose
the identity component $\Ctilde nt$, we can eliminate the zero-forms
$\eta$. Thus, we can represent the contributions to
our v.e.v.'s graphically as follows:
we represent the knot as a horizontal line (which we suppose
directed from left to right) with the base point on its boundary
and the spanning surface as the portion of plane above it, and the
two-forms $\theta_{ij}$ as arrows connecting the point $i$
to the point $j$ (s.\ fig.\ \ref{fig-examples}).

We can give even a better 
description---cfr.\ \cite{BT}---if we introduce the bundles
\beq
\begin{array}{c}
\Ctilde nt(S^1\times\calP\times\calS)\\
\Big\downarrow\vcenter{\rlap{$p$}}\\
S^1\times\calP\times\calS
\end{array},
\eeq
where $\calP$ is the space of the maps $S^1\to S^1$ homotopic to
the identity and $\calS$ is the space of imbeddings of ${B^2}$ in
${\reali^3}$, and the fiber is $\Ctilde nt$.
Now we can see a generic v.e.v.\ as the sum of 
contributions which read
\beq
I = p_* [I],
\eeq
where $p_*$ denotes the push forward along the fiber,
and $[I]$ is a form on the bundle.

A locally constant function on the base
space---i.e., a function whose differential vanishes)---will 
be a topological invariant on $\calS$, for
a locally constant function is a constant
on $S^1\times\calP$ which are connected.
If, moreover, this topological invariant does not depend
on cohomological deformations of the surface, it will 
eventually be an invariant of its boundary.

As in \cite{BT}, the differential of $I$
can be written, by Stokes's theorem, as
\beq
dI = p_*d[I] + p_*^\de[I] = p_*^\de[I], 
\lab{dI}
\eeq
where $p_*^\de$ denotes the push forward along the boundary of 
the fiber.

It is useful to distinguish on this boundary between
{\em principal} and {\em hidden} faces. \label{simple}
The principal faces are essentially of four types:
\begin{enumerate}
\item two points on the knot collapse together;
\item one point on the knot collapses to the base point;
\item two point on the surface collapse together;
\item one point on the surface collapses on the knot where either
{\em (a)} there are no points, or {\em (b)} 
there is one point, or {\em (c)} there is the base point.
\end{enumerate}
All the other components of the boundary (viz., when more points
come together) are referred to as the hidden faces.
Among the principal faces, we will call {\em simple} 
the ones of type 1, 2 and 3a.

The principal-face contribution $\delta I$ to $dI$ can be evaluated
``graphically'' just by looking at the diagrams.
What is not immediate is seeing wether the push forward along the
hidden faces vanishes. However, in App.\ \ref{app-hf} we prove a vanishing
theorem for the push forward along all faces but the simple principal faces 
(s. Thm.\ \ref{th-vt}).

\subsection{The perturbative expansion of $\ave{\calO_0}$}
By \rf{setstar}, the gauge-fixed observable $\gammatot$ \rf{gammatot}
now reads
\beq
\gammatotp\gf = \gamma^{\Sigma_K}_\gf + \gamma^{(K,\xnot)}_\gf +
\gamma^{(\Sigma_K,\xnot)}_\gf,
\lab{gammatotgf}
\eeq
with
\beqy
\gamma^{\Sigma_K}_\gf &=&
\intsig [B\wedge A + (*d\bar\psi)\psi + c\,(*d\bar c),
\lab{gammasigmagf}\\
\gamma^{(K,\xnot)}_\gf  &=&
\intcdue A_1\eta_{12}\wedge B_2,\lab{gammaKgf}\\
\gamma^{(\Sigma_K,\xnot)}_\gf &=&
\psi(\xnot)\,\intsig (*d\bar\psi) + 
\intsig (*d\bar c)\,c(\xnot).\lab{gamma0gf}
\eeqy

\subsubsection{The general structure of the perturbative
expansion}
\lab{ssec-gspe}
The first thing we notice is that all these functionals are odd
under \rf{parity}. Thus, only an even product of them will have a
nonvanishing v.e.v. This, by the way, proves that 
$\ave{\calO_0}_\lambda$ is even in $\lambda$, in according with
\rf{aveOK0}.

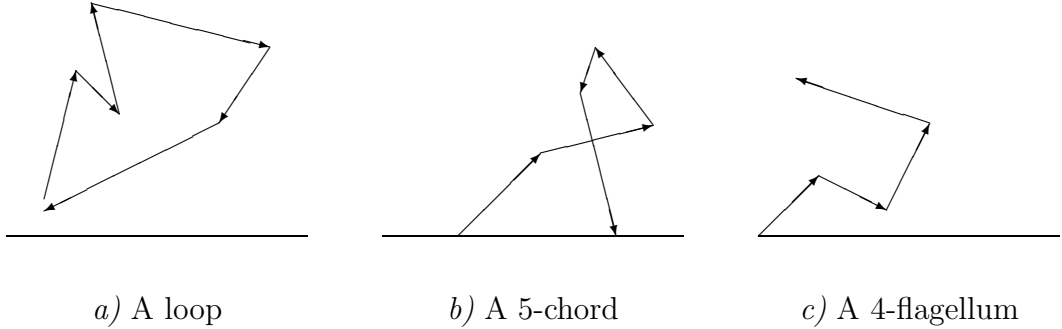
\begin{figure}
\unitlength 1.00mm
\linethickness{0.4pt}
\begin{picture}(140.00,51.01)
\put(0.00,20.00){\line(1,0){40.00}}
\put(5.00,25.00){\vector(1,4){4.25}}
\put(9.25,42.00){\vector(1,-1){5.75}}
\put(15.00,36.25){\vector(-1,4){3.69}}
\put(11.31,51.00){\vector(4,-1){23.69}}
\put(35.00,45.08){\vector(-2,-3){6.72}}
\put(28.28,35.00){\vector(-2,-1){23.28}}
\put(50.00,20.00){\line(1,0){40.00}}
\put(100.00,20.00){\line(1,0){40.00}}
\put(60.00,20.00){\vector(1,1){11.00}}
\put(71.00,31.00){\vector(4,1){15.00}}
\put(86.00,34.75){\vector(-3,4){7.69}}
\put(78.31,45.00){\vector(-1,-3){2.00}}
\put(76.31,39.00){\vector(1,-4){4.75}}
\put(100.00,20.00){\vector(1,1){8.00}}
\put(108.00,28.00){\vector(2,-1){9.00}}
\put(117.00,23.50){\vector(1,2){5.75}}
\put(122.75,35.00){\vector(-3,1){17.75}}
\put(20.00,10.00){\makebox(0,0)[cc]{{\em a)} A loop}}
\put(70.00,10.00){\makebox(0,0)[cc]{{\em b)} A $5$-chord}}
\put(120.00,10.00){\makebox(0,0)[cc]{{\em c)} A $4$-flagellum}}
\end{picture}
\caption{Elements that appear in the diagrams}\label{fig-ele}
\end{figure}

The second observation is that, by \rf{defQ},
the functional $\gamma^{\Sigma_K}_\gf$, is  $Q$-exact, viz.,
\beq
\gamma^{\Sigma_K}_\gf = \intsig dx^\alpha\wedge dx^\beta
\,[Q(\psi A - Bc)]_{\alpha\beta};
\eeq
thus, the v.e.v.\ of any of its powers vanishes. This implies that
no loops appear among the v.e.v.'s we are computing.
In fig.\ \ref{fig-ele}.a , we show one of these loops.
Notice that, if we were working on a less trivial manifold, \rf{Sgf}
would not be supersymmetric, so such loops would exist. As a matter of
fact, the v.e.v.\ considered in \rf{aveOSigma} consists entirely of
such loop diagrams.

The third observation is that $\gamma^{(\Sigma_K,\xnot)}_\gf$ is
linear in the Grassmann variables $c(\xnot)$ and $\psi(\xnot)$.
Thus, its square simply reads
\beq
(\gamma^{(\Sigma_K,\xnot)}_\gf)^2= 
\psi(\xnot)\,\intsig (*d\bar\psi)\ \intsig (*d\bar c)\,c(\xnot),
\eeq
while all higher powers vanish.

We can now describe the features of the perturbative expansion
of $\ave{\calO_0}_\lambda$. In the v.e.v.'s where no 
$\gamma^{(\Sigma_K,\xnot)}_\gf$ appears, we have to Wick contract
the fields $A$ and $B$ on $K$ and on $\Sigma_K$ together in all
possible ways discarding all the diagrams that contain a loop.
Thus, we are left with chains that connect two points on $K$ through
a certain number of bivalent vertices
on $\Sigma_K$. We will call an {\em $n$-chord} such a chain, where
$n>0$ is the number of links (s.\ fig.\ \ref{fig-ele}.b).

If the v.e.v.\ contains $\gamma^{(\Sigma_K,\xnot)}_\gf$,
besides the chords described above, 
we will have a chain connecting $\xnot$ with a point on $\Sigma_K$ 
through a certain number of bivalent vertices on $\Sigma_K$. 
We will call an {\em $n$-flagellum} such a chain 
(s.\ fig.\ \ref{fig-ele}.c).
 
Eventually, we have v.e.v.'s containing 
$(\gamma^{(\Sigma_K,\xnot)}_\gf)^2$. They contain some chords and
two flagella. 

For a v.e.v. not to vanish, the total number of links in
all the chords must be even; moreover,
the total number of links in the flagella
must be even.

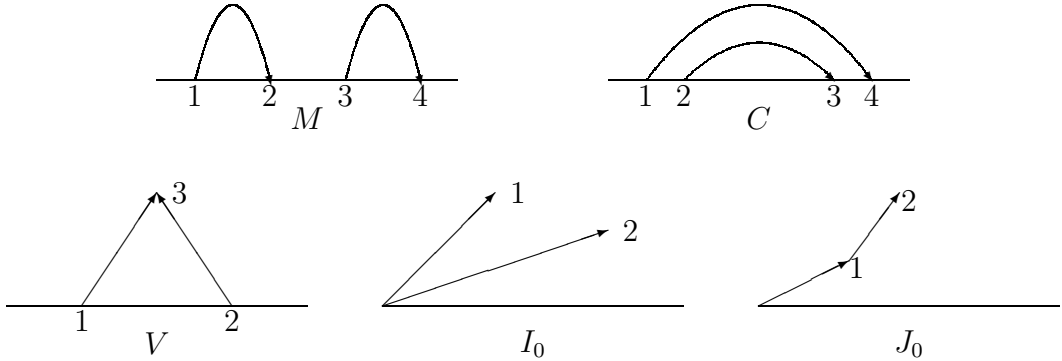
\begin{figure}
\unitlength 1.00mm
\linethickness{0.4pt}
\begin{picture}(140.00,60.00)
\put(20.00,40.00){\line(1,0){40.00}}
\put(80.00,40.00){\line(1,0){40.00}}
%\bezvec{164}(25.00,40.00)(30.00,60.00)(35.00,40.00)
\put(35.00,40.00){\vector(1,-4){0.2}}
\bezier{164}(25.00,40.00)(30.00,60.00)(35.00,40.00)
%\end
%\bezvec{164}(45.00,40.00)(50.00,60.00)(55.00,40.00)
\put(55.00,40.00){\vector(1,-4){0.2}}
\bezier{164}(45.00,40.00)(50.00,60.00)(55.00,40.00)
%\end
%\bezvec{200}(85.00,40.00)(100.00,60.00)(115.00,40.00)
\put(115.00,40.00){\vector(3,-4){0.2}}
\bezier{200}(85.00,40.00)(100.00,60.00)(115.00,40.00)
%\end
%\bezvec{112}(90.00,40.00)(100.00,50.00)(110.00,40.00)
\put(110.00,40.00){\vector(1,-1){0.2}}
\bezier{112}(90.00,40.00)(100.00,50.00)(110.00,40.00)
%\end
\put(40.00,35.00){\makebox(0,0)[cc]{$M$}}
\put(100.00,35.00){\makebox(0,0)[cc]{$C$}}
\put(0.00,10.00){\line(1,0){40.00}}
\put(50.00,10.00){\line(1,0){40.00}}
\put(100.00,10.00){\line(1,0){40.00}}
\put(10.00,10.00){\vector(2,3){10.00}}
\put(30.00,10.00){\vector(-2,3){10.00}}
\put(50.00,10.00){\vector(1,1){15.00}}
\put(50.00,10.00){\vector(3,1){30.00}}
\put(100.00,10.00){\vector(2,1){12.00}}
\put(112.00,16.00){\vector(3,4){6.75}}
\put(20.00,5.00){\makebox(0,0)[cc]{$V$}}
\put(70.00,5.00){\makebox(0,0)[cc]{$I_0$}}
\put(120.00,5.00){\makebox(0,0)[cc]{$J_0$}}
\put(25.00,38.00){\makebox(0,0)[cc]{$1$}}
\put(35.00,38.00){\makebox(0,0)[cc]{$2$}}
\put(45.00,38.00){\makebox(0,0)[cc]{$3$}}
\put(55.00,38.00){\makebox(0,0)[cc]{$4$}}
\put(85.00,38.00){\makebox(0,0)[cc]{$1$}}
\put(90.00,38.00){\makebox(0,0)[cc]{$2$}}
\put(110.00,38.00){\makebox(0,0)[cc]{$3$}}
\put(115.00,38.00){\makebox(0,0)[cc]{$4$}}
\put(10.00,8.00){\makebox(0,0)[cc]{$1$}}
\put(30.00,8.00){\makebox(0,0)[cc]{$2$}}
\put(23.00,25.00){\makebox(0,0)[cc]{$3$}}
\put(68.00,25.00){\makebox(0,0)[cc]{$1$}}
\put(83.00,20.00){\makebox(0,0)[cc]{$2$}}
\put(113.00,15.00){\makebox(0,0)[cc]{$1$}}
\put(120.00,24.00){\makebox(0,0)[cc]{$2$}}
\end{picture}
\caption{The second-order diagrams}\label{fig-MCVIJ}
\end{figure}

In figs.\ \ref{fig-examples} and \ref{fig-MCVIJ}, 
some diagrams of nonvanishing v.e.v.'s are shown. 
At order $2n$ in $\lambda$, we will have a sum of diagrams of
this kind with a total number of links equal to
$2n$.

From the structure of the perturbative expansion, it is easy to see
that, if $\avel{\calO_0}$
is a topological invariant, then it is invariant under $\Sigma_K\to
\Sigma_K+\de T$ with $T\in\Omega_3(\reali^3)$. In fact, if
we have a topological invariant, we can move the region where this
deformation occurs to infinity. Since all the vertices on $\Sigma_K$
are connected through a finite number of links to a point on $K$,
this region at infinity does not contribute. This, of course, would not
be true if loops on $\Sigma_K$ were allowed.

As a final remark, we notice that, if we represent the unknot
in standard framing as a planar curve and choose its spanning
surface to belong to the same plane, then
\beq
\ave{\calO_0[\bigcirc_\sfram,\lambda]}=1.
\eeq

\subsubsection{The second order}
Now we want to compute explictly the v.e.v. of 
$\frac12(\gammatot)^2$.
By the supersymmetry argument, we know that the v.e.v.\ of
$(\gammaSK)^2$ vanishes. Moreover, the v.e.v.\ of 
$\gammaKnot\gammaSnot$ is equal to the product of the v.e.v.'s
of $\gammaKnot$ and $\gammaSnot$, which vanish. Thus, we are left
with only four contributions, which, after some computations,
can be written as
\beq
\begin{array}{lcl}
\avel{\frac12 (\gammaKnot)^2} &=& \lambda^2\,(M-C),\\
\avel{\gammaKnot\gammaSK} &=& \lambda^2\, V,\\
\avel{\frac12(\gammaSnot)^2} &=& \lambda^2\, I_0,\\
\avel{\gammaSK\gammaSnot} &=& -2\lambda^2 J_0,
\end{array}
\eeq
where the diagrams $M$, $C$, $V$, $I_0$ and $J_0$ are shown
in fig.\ \ref{fig-MCVIJ}. Explicitly they read
\beq
\begin{array}{lcl}
M &=& \int_{\Ctilde 40}\theta_{12} \wedge \theta_{34},\\
C &=& \int_{\Ctilde 40}\theta_{14} \wedge \theta_{23},\\
V &=& \int_{\Ctilde 21}\theta_{13} \wedge \theta_{23},\\
I_0 &=& \int_{\Ctilde 02}\theta_{01} \wedge \theta_{02},\\
J_0 &=& \int_{\Ctilde 02}\theta_{01} \wedge \theta_{12}.
\end{array}
\eeq
Thus, the second order of $\avel{\calO_0}$ reads
\beq
\avel{\frac12(\gammatot)^2} = \lambda^2\,(M-C+V+I_0-2J_0).
\lab{avegamma2}
\eeq

\subsection{The v.e.v.\ of the corrected observable}
As explained in subsection \ref{ssec-spko}, we do not expect
$\avel{\calO_0}$ to be a knot invariant (at least not with a general
framing), since, in general, $\calO_0$ is {\em not} an observable.
In subsection \ref{ssec-cspko}, we have seen that there is a 
procedure that leads to an observable $\calO$ starting from $\calO_0$.
We have computed the first correction \rf{O1} explicitly and
have shown that the corrected second-order v.e.v.\ is given
by the v.e.v.\ of the observable $O_2$ defined in \rf{O2}.
In this section, we will compute this v.e.v.\ explicitly and show that
it is a knot invariant.

\subsubsection{The v.e.v. of $O_2$}
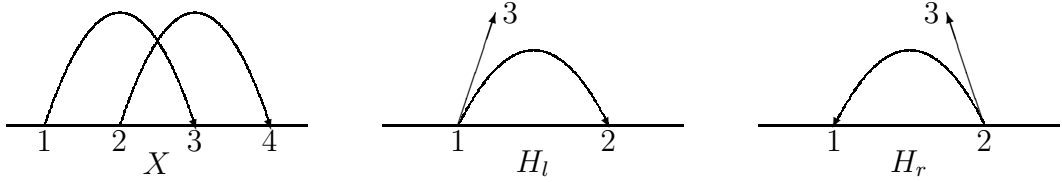
\begin{figure}
\unitlength 1.00mm
\linethickness{0.4pt}
\begin{picture}(140.00,40.00)
\put(50.00,10.00){\line(1,0){40.00}}
%\bezvec{180}(60.00,10.00)(70.00,30.00)(80.00,10.00)
\put(80.00,10.00){\vector(1,-2){0.2}}
\bezier{180}(60.00,10.00)(70.00,30.00)(80.00,10.00)
%\end
\put(60.00,10.00){\vector(1,3){5.00}}
\put(70.00,5.00){\makebox(0,0)[cc]{$H_l$}}
\put(140.00,10.00){\line(-1,0){40.00}}
%\bezvec{180}(130.00,10.00)(120.00,30.00)(110.00,10.00)
\put(110.00,10.00){\vector(-1,-2){0.2}}
\bezier{180}(130.00,10.00)(120.00,30.00)(110.00,10.00)
%\end
\put(130.00,10.00){\vector(-1,3){5.00}}
\put(120.00,5.00){\makebox(0,0)[cc]{$H_r$}}
\put(0.00,10.00){\line(1,0){40.00}}
%\bezvec{252}(5.00,10.00)(15.00,40.00)(25.00,10.00)
\put(25.00,10.00){\vector(1,-3){0.2}}
\bezier{252}(5.00,10.00)(15.00,40.00)(25.00,10.00)
%\end
%\bezvec{252}(15.00,10.00)(25.00,40.00)(35.00,10.00)
\put(35.00,10.00){\vector(1,-3){0.2}}
\bezier{252}(15.00,10.00)(25.00,40.00)(35.00,10.00)
%\end
\put(20.00,5.00){\makebox(0,0)[cc]{$X$}}
\put(5.00,8.00){\makebox(0,0)[cc]{$1$}}
\put(15.00,8.00){\makebox(0,0)[cc]{$2$}}
\put(25.00,8.00){\makebox(0,0)[cc]{$3$}}
\put(35.00,8.00){\makebox(0,0)[cc]{$4$}}
\put(60.00,8.00){\makebox(0,0)[cc]{$1$}}
\put(80.00,8.00){\makebox(0,0)[cc]{$2$}}
\put(110.00,8.00){\makebox(0,0)[cc]{$1$}}
\put(130.00,8.00){\makebox(0,0)[cc]{$2$}}
\put(67.00,25.00){\makebox(0,0)[cc]{$3$}}
\put(123.00,25.00){\makebox(0,0)[cc]{$3$}}
\end{picture}
\caption{The second-order correction diagrams}\label{fig-XHH}
\end{figure}

To evaluate the v.e.v.\ of the correction $i\lambda U_1$, we notice
that, by \rf{Uu} and Statement \ref{statBV},
\beq
\avel{i\lambda U_1} = \avel{i\lambda\Delta u_1} =
\avel{\sigma u_1}. 
\eeq
By \rf{u1}, we have then
\beq
\avel{i\lambda U_1} = \tildeU_1 + \tildeU_2,
\eeq
with
\beq
\begin{array}{lcl}
\tildeU_1 &=& 
\avel{\gamma_{ABB}\,\sigma\intsig B^*
+\gamma_{BAA}\,\sigma\intsig A^*},\\
\tildeU_2 &=& -
\avel{(\sigma\gamma_{ABB})\,\intsig B^*
+(\sigma\gamma_{BAA})\,\intsig A^*}.
\end{array}
\eeq
Finally, an explicit evaluation of this v.e.v.'s yields
\beq
\begin{array}{lcl}
\tildeU_1 &=& 2\lambda^2\,(X-2M-C),\\
\tildeU_2 &=& \lambda^2\,(H_l+H_r),
\end{array}
\eeq
where the new diagrams $X$, $H_l$ and $H_r$ are shown in 
fig.\ \ref{fig-XHH}.
Explicitly they read
\beq
\begin{array}{lcl}
X &=& \int_{\Ctilde 40} \theta_{13}\wedge\theta_{24},\\
H_l &=& \int_{\Ctilde 21} \theta_{12}\wedge\theta_{13},\\
H_r &=& \int_{\Ctilde 21} \theta_{21}\wedge\theta_{23}.
\end{array}
\eeq
Therefore, the correction to $\avel{\calO_0}$ reads
\beq
\avel{i\lambda U_1}=\lambda^2\,(2X-4M-2C+H_l+H_r).
\lab{aveU1}
\eeq
To get the complete v.e.v.\ of $O_2$, we have to add 
\rf{aveU1} to \rf{avegamma2}. 
First, however, it is useful to notice that the square
of the self-linking number
\beq
\slk K = 2\int_{\Ctilde 20} \theta_{12}
\eeq
can be written as
\beq
(\slk K)^2 = 8(M+C-X);
\eeq
thus,
\beq
\avel{O_2} = -\frac38\,\lambda^2 (\slk K)^2  
+ \lambda^2\, w_2,
\lab{aveO2}
\eeq
with
\beq
w_2 = -X+V+H_l+H_r+I_0-2J_0.
\lab{defw2}
\eeq
We conclude this subsection by noticing that, if we take
the unknot as a planar curve and choose its spanning surface
to lie in the same surface, it is immediately proved that
\beq
w_2(\bigcirc)=0.
\lab{w200}
\eeq
Shortly we will prove that $w_2$ is a knot
invariant; thus, \rf{w200} actually holds for any presentation of the
unknot.

\subsubsection{The invariance of the second-order term}
\lab{ssec-isot}
Now we will show that the principal-face variation of
$w_2$ vanishes; we will follow the approach of Ref.\ \cite{BT},
which we have recalled in subsection \ref{ssec-reg}.

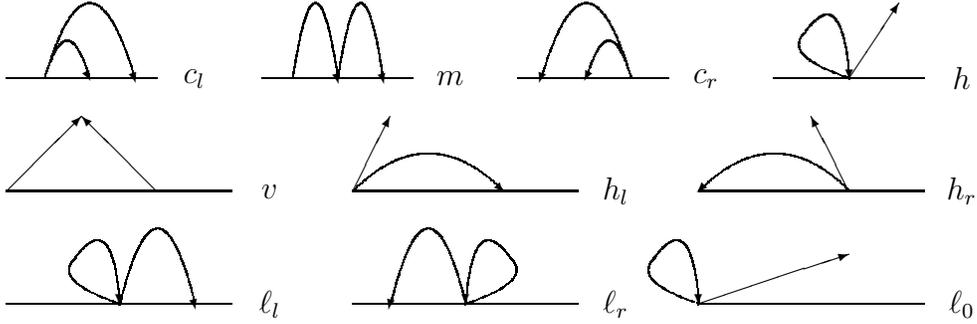
\begin{figure}
\unitlength 1.00mm
\linethickness{0.4pt}
\begin{picture}(127.00,60.00)
\put(0.00,10.00){\line(1,0){30.00}}
\put(92.00,10.00){\line(1,0){30.00}}
%\bezvec{164}(15.00,10.00)(20.00,30.00)(25.00,10.00)
\put(25.00,10.00){\vector(1,-4){0.2}}
\bezier{164}(15.00,10.00)(20.00,30.00)(25.00,10.00)
%\end
\bezier{56}(15.00,10.00)(6.00,13.00)(9.00,16.00)
%\bezvec{88}(9.00,16.00)(14.00,23.00)(15.00,10.00)
\put(15.00,10.00){\vector(0,-1){0.2}}
\bezier{88}(9.00,16.00)(14.00,23.00)(15.00,10.00)
%\end
\put(76.00,10.00){\line(-1,0){30.00}}
%\bezvec{164}(61.00,10.00)(56.00,30.00)(51.00,10.00)
\put(51.00,10.00){\vector(-1,-4){0.2}}
\bezier{164}(61.00,10.00)(56.00,30.00)(51.00,10.00)
%\end
\bezier{56}(61.00,10.00)(70.00,13.00)(67.00,16.00)
%\bezvec{88}(67.00,16.00)(62.00,23.00)(61.00,10.00)
\put(61.00,10.00){\vector(0,-1){0.2}}
\bezier{88}(67.00,16.00)(62.00,23.00)(61.00,10.00)
%\end
\bezier{56}(92.00,10.00)(83.00,13.00)(86.00,16.00)
%\bezvec{88}(86.00,16.00)(91.00,23.00)(92.00,10.00)
\put(92.00,10.00){\vector(0,-1){0.2}}
\bezier{88}(86.00,16.00)(91.00,23.00)(92.00,10.00)
%\end
\put(92.00,10.00){\vector(3,1){20.00}}
\put(35.00,10.00){\makebox(0,0)[cc]{$\ell_l$}}
\put(81.00,10.00){\makebox(0,0)[cc]{$\ell_r$}}
\put(127.00,10.00){\makebox(0,0)[cc]{$\ell_0$}}
\put(0.00,25.00){\line(1,0){30.00}}
\put(46.00,25.00){\line(1,0){30.00}}
\put(0.00,25.00){\vector(1,1){10.00}}
\put(20.00,25.00){\vector(-1,1){10.00}}
%\bezvec{112}(46.00,25.00)(56.00,35.00)(66.00,25.00)
\put(66.00,25.00){\vector(1,-1){0.2}}
\bezier{112}(46.00,25.00)(56.00,35.00)(66.00,25.00)
%\end
\put(46.00,25.00){\vector(1,2){5.00}}
%\bezvec{112}(112.00,25.00)(102.00,35.00)(92.00,25.00)
\put(92.00,25.00){\vector(-1,-1){0.2}}
\bezier{112}(112.00,25.00)(102.00,35.00)(92.00,25.00)
%\end
\put(112.00,25.00){\vector(-1,2){5.00}}
\put(92.00,25.00){\line(1,0){30.00}}
\put(35.00,25.00){\makebox(0,0)[cc]{$v$}}
\put(81.00,25.00){\makebox(0,0)[cc]{$h_l$}}
\put(127.00,25.00){\makebox(0,0)[cc]{$h_r$}}
\put(0.00,40.00){\line(1,0){20.00}}
%\bezvec{84}(5.00,40.00)(8.00,50.00)(11.00,40.00)
\put(11.00,40.00){\vector(1,-2){0.2}}
\bezier{84}(5.00,40.00)(8.00,50.00)(11.00,40.00)
%\end
%\bezvec{168}(5.00,40.00)(11.00,60.00)(17.00,40.00)
\put(17.00,40.00){\vector(1,-3){0.2}}
\bezier{168}(5.00,40.00)(11.00,60.00)(17.00,40.00)
%\end
\put(88.00,40.00){\line(-1,0){20.00}}
%\bezvec{84}(83.00,40.00)(80.00,50.00)(77.00,40.00)
\put(77.00,40.00){\vector(-1,-2){0.2}}
\bezier{84}(83.00,40.00)(80.00,50.00)(77.00,40.00)
%\end
%\bezvec{168}(83.00,40.00)(77.00,60.00)(71.00,40.00)
\put(71.00,40.00){\vector(-1,-3){0.2}}
\bezier{168}(83.00,40.00)(77.00,60.00)(71.00,40.00)
%\end
\put(34.00,40.00){\line(1,0){20.00}}
%\bezvec{160}(38.00,40.00)(41.00,60.00)(44.00,40.00)
\put(44.00,40.00){\vector(1,-4){0.2}}
\bezier{160}(38.00,40.00)(41.00,60.00)(44.00,40.00)
%\end
%\bezvec{160}(44.00,40.00)(47.00,60.00)(50.00,40.00)
\put(50.00,40.00){\vector(1,-4){0.2}}
\bezier{160}(44.00,40.00)(47.00,60.00)(50.00,40.00)
%\end
\bezier{56}(112.00,40.00)(103.00,43.00)(106.00,46.00)
%\bezvec{88}(106.00,46.00)(111.00,53.00)(112.00,40.00)
\put(112.00,40.00){\vector(0,-1){0.2}}
\bezier{88}(106.00,46.00)(111.00,53.00)(112.00,40.00)
%\end
\put(102.00,40.00){\line(1,0){20.00}}
\put(112.00,40.00){\vector(2,3){6.67}}
\put(25.00,40.00){\makebox(0,0)[cc]{$c_l$}}
\put(93.00,40.00){\makebox(0,0)[cc]{$c_r$}}
\put(59.00,40.00){\makebox(0,0)[cc]{$m$}}
\put(127.00,40.00){\makebox(0,0)[cc]{$h$}}
\end{picture}
\caption{The principal-face diagrams}\label{fig-pf}
\end{figure}

Referring to figs.\ \ref{fig-MCVIJ}, \ref{fig-XHH} and
\ref{fig-pf},
we start considering the principal-face contributions
\beq
\begin{array}{lcl}
\delta X &=& c_l - m + c_r,\\
\delta V &=& c_l + m + c_r - 2v + \ell_l + \ell_r,\\
\delta H_l &=& -m + h - h_l + h_r -\ell_l,\\
\delta H_r &=& -m - h - h_l + h_r -\ell_r,\\
\delta I_0 &=& 2h_l + 2\ell_0,\\
\delta J_0 &=& -v + h_r + \ell_0,
\end{array}
\lab{delta...}
\eeq
which by \rf{defw2} imply
\beq
\delta w_2 = 0.
\lab{deltaw2}
\eeq

It should be clear from fig.\ \ref{fig-pf} what $c_l$, $m$, $c_r$,
$v$, $h_l$ and $h_r$ mean. To write the diagrams
$h$, $\ell_l$, $\ell_0$ and $\ell_r$, we need to introduce
explicitly the map
\beq
\Phi:B^2\longrightarrow\reali^3
\eeq
that defines the surface $\Sigma_K$.
The diagram $h$ is given by
\beq
h = \int_{\Ctilde 11} \theta_{12}\wedge\theta_{11},
\eeq
where $\theta_{11}$ is the pull back of the volume form $\omega$
through the map
\beq
\phi_{11}(\vec x) = \frac{\dot\Phi(x_1)}{|\dot\Phi(x_1)|},
\quad \vec x\in\Ctilde 11.
\eeq
and $\dot\Phi$ denotes the derivative of $\Phi$ in the direction
tangent to the knot (notice that $\Phi(x_1)$ is on the knot).
To describe the remaining diagrams, we have also to introduce 
$\Phi'$, i.e., the derivative of $\Phi$ w.r.t.\ the other coordinate in 
the parametrization of the surface. In general the vector $\Phi'$ will 
{\em not} be orthogonal to $\dot\Phi$. To obtain an orthogonal vector
we define
\beq
\Phior = \Phi' - \frac{\Phi'\cdot\dot\Phi}{|\dot\Phi|^2}\dot\Phi.
\lab{defPhiortho}
\eeq
Then the diagrams $\ell_l$ and $\ell_r$ read
\beq
\begin{array}{lcl}
\ell_l = \int_{\Ctilde 20} [\theta_{12}\wedge
\int_{\calU_2}\theta_2],\\
\ell_r = \int_{\Ctilde 20} [\theta_{12}\wedge
\int_{\calU_1}\theta_1],
\end{array}
\eeq
where the one-dimensional manifolds $\calU_i$ are defined as
\beq
\begin{array}{ll}
\calU_i=\{(u_1,u_2^1,u_2^2)\in\reali\times\reali\times\reali^+/
&[(u_1)^2 + (u_2^1)^2]|\dot\Phi(x_i)|^2+(u_2^2)^2|\Phior(x_i)|^2=1,\\
& u_1 + u_2^1 = 0\},
\end{array}
\eeq
and $\theta_i$ is the pull back of $\omega$ to $\calU_i$ through
the map
\beq
\phi_i(u_1,u_2^1,u_2^2) = \frac
{(u_2^1-u_1)\dot\Phi(x_i)+u_2^2\Phior(x_i)}
{|(u_2^1-u_1)\dot\Phi(x_i)+u_2^2\Phior(x_i)|}.
\eeq
Finally,
\beq
\ell_0 = \int_{\Ctilde 01} [\theta_{01}\wedge
\int_{\widetilde\calU_0}\widetilde\theta_0],
\eeq
where the one-dimensional manifold $\widetilde\calU_0$ is defined as
\beq
\widetilde\calU_0 = \{(u^1,u^2)\in\reali\times\reali^+/
(u^1)^2|\dot\Phi(\xnot)|^2 + (u^2)^2|\Phior(\xnot)|^2 = 1\},
\eeq
and $\widetilde\theta_0$ is the pull back of $\omega$ 
to $\widetilde\calU_0$ through the map
\beq
\widetilde\phi_0(u_1,u_2)=\frac
{u^1\dot\Phi(\xnot) + u^2\Phior(\xnot)}
{|u^1\dot\Phi(\xnot) + u^2\Phior(\xnot)|}.
\eeq

Notice that in \rf{delta...} we have written only the non vanishing
contributions. (Actually, more sophisticated arguments, s.\ App.\ 
\ref{app-hf}, show that also $\ell_l$, $\ell_r$ and $\ell_0$ vanish.)
\ \ All other possible terms vanish for one of
the following reasons:
\begin{enumerate}
\item we have to integrate a form on a space of lower dimension;
\item a factor $\theta_{ij}^2$ appears, or
\item the push forward vanish because of a symmetry.
\end{enumerate}
An example of the first case is the push forward of $[V]$ along
the face obtained by sending $3$ to $0$ which gives
\[
\int_{\widetilde\calU_0}[\int_{\Ctilde 20} \theta_{10}\wedge\theta_{20}].
\]
The second case happens, e.g., when we push forward $[V]$
along the face where we send $1$ to $2$.
The third case occurs in the push forward of $[J_0]$ when we send $1$
to $2$, the symmetry being the exchange of $1$ with $2$ which does
not reverse the orientation of the manifold we are integrating over but
changes the sign of the form to be integrated.

For the same reasons, the push forwards 
of $[X]$, $[I_0]$, $[J_0]$ and $[H_l]+[H_r]$
along the hidden faces vanish. 
The only non-trivial case is the push forward of $[V]$ 
along the hidden face where $1$, $2$ and $3$ come together.
This case is analyzed in App.\ \ref{app-hf}, and a vanishing theorem
is proved.

These results together with \rf{deltaw2} prove the following
\begin{Th}
The corrected second-order term $w_2$ is a topological invariant
of the imbedding $\Sigma_K$ of $B^2$ in $\reali^3$.
\lab{thm-Sigmainv}
\end{Th}
As a consequence, if we deform the imbedding 
$\Sigma_K$ by adding to it the boundary of a three-cycle,
we can always move this deformation to infinity. Since all the vertices
on $\Sigma_K$ are connected through at most two $\theta$'s to a point
living on $K$, the deformation at infinity will not contribute.
Therefore, $w_2$ actually depends only on $K$, and we have the
following
\begin{Th}
The corrected second-order term $w_2$ is a knot invariant.
\lab{thm-Kinv}
\end{Th}
The chord-diagram contribution $-X$ to $w_2$ is exactly the same
that appears in the invariant studied in \cite{GMM,BN-th}.
This invariant is known to be equal to the second coefficient $a_2$ of 
the Alexander--Conway polynomial plus a constant term (viz., the
value it takes on the unknot).

Notice that the chord diagram $-X$ alone is {\em not} a knot invariant.
To get a knot invariant we have to add to it
either the other terms that define
$w_2$---let call $W$ their sum---or the diagram $Y$ considered in
\cite{GMM,BN-th,BT} (viz., a diagram with a trivalent vertex in 
$\reali^3$).
Since both $-X+W$ and $-X+Y$ are knot invariants, also $T=W-Y$ is a knot 
invariant. Our claim is that $T$ is trivial (i.e., it is the 
same for all knots). To prove it, it is enough to check that $T$ takes
the same value on two knots $K_+$ and $K_-$ that differ only around a 
chosen crossing. We notice that the difference 
$T(K_+)-T(K_-)$ comes from a singularity at the crossing point 
where the flip occurs---as in \cite{BN-th}---or along the line 
where the two spanning surfaces get to intersect. 
However, it is not difficult to check that such singularities do not arise; so $T$ is a constant.

Therefore, $w_2$ is equal to $a_2$ plus a constant.
However, since by definition $a_2(\bigcirc)=0$, \rf{w200} implies
\beq
w_2 = a_2,
\eeq
and Conjecture \fullref{conj-KAC} is satisfied at this order.

As a concluding remark, we notice that in passing from $Y$ to $W$
one of the integrations on the knot is replaced by an integration
on the spanning surface; so it should be possible
to relate $W$ and $Y$ directly via Stokes's theorem.

\subsubsection{Higher orders}
Thm.\ \ref{th-notan} ensures that a quantum observable $\calO$ extending
$\calO_0$ exists. Its v.e.v.\ at order $\lambda^n$ will be given by 
diagrams
containing $n$ propagators connecting points on the knot and/or on the
spanning surface.  Of course, the restrictions given in subsection 
\ref{ssec-gspe} for the v.e.v.\ of $\calO_0$ do not hold anymore;
in particular, the vertices on the knot will not necessarily be 
univalent and the vertices on the spanning surface will not necessarily
be bivalent (we have already seen a counterexample at the second order).
However, no loops on the surface will appear (since the corrections must
vanish when the spanning surface is boundariless). Moreover, since the
v.e.v.\ of $\calO_0$ vanishes at odd order, we do not need odd-order
corrections.

The combinatorics of these diagrams will be dictated by the specific 
form of the corrected observable $\calO$. 
What we expect, by field-theoretical arguments, is that these 
combinations of diagrams will be metric independent, i.e., will
be the sum of invariants possibly times powers of the self-linking 
number, i.e., ``isolated chords" in the diagrams. The true invariants 
will then be obtained by factorizing the isolated chords.

A rigorous mathematical proof that they are actually knot invariants
will simply require checking that the
principal-face contributions of the diagrams that sum up cancel each 
other, for Thm.\ \fullref{th-vt} ensures that the push forwards along 
hidden faces always vanish.

Notice that now we could also throw away the $BF$ field theory and
directly study $\delta$-closed combinations of diagrams with vertices
on the knot and/or on the spanning surface.  By Thm.\ \ref{th-vt}, these
will yield knot invariants as well as higher-degree cohomology classes
on the space of imbeddings (the degree being given by $2l-n-2t$ where 
$l$ is the number of propagators, $n$ the number of points on the 
knot and $t$ the number of points on the surface).

\section{A glimpse to higher dimensions}
\lab{sec-glim}
There is no problem in defining the Abelian $BF$ theory in any dimension:
just take $A$ and $B$ to be fields taking values in $\Omega^p(M)$
and $\Omega^q(M)$ respectively, with $p+q+1=d$ and $d=\dim M$.
The classical action \rf{Sclomega} can easily be extended
to a BV action. The partition function then is known to be
equal to the Ray--Singer torsion or to its inverse (depending on $p$)
\cite{Schw,BlT}. 
M
oreover, it is not difficult to generalize
the observable \rf{defgammaSigma}, where now $\Sigma\in H_{d-1}(M,\de M)$.
The classical part of this observable reads
\beq
\gamma^\Sigma_\cl = \int_\Sigma B\wedge A,
\eeq
and satisfies
\beq
s\gamma^\Sigma_\cl = 0,\quad\mbox{on shell},
\eeq
where {\em on shell} means modulo the classical equations of motion,
\beq
dA=0,\quad dB=0,
\eeq
and $s$ is the BRST operator
\beq
sA=dc,\quad sB=d\psi
\eeq
(now $c$ and $\psi$ are a $(p-1)$- and a $(q-1)$-form respectively).

If $\Sigma_K$ is a spanning surface for a $(d-2)$-knot $K$
(i.e., an imbedding of $S^{d-2}$ in $\reali^d$), then
\beq
s\gamma^{\Sigma_K}_\cl = \oint_K[\psi\wedge A + (-1)^q\, B\wedge c],
\quad\mbox{on shell}.
\lab{sgSd}
\eeq
To get an on-shell $s$-closed functional, we have to add to
$\gamma^{\Sigma_K}$ another term canceling the r.h.s.\ of
\rf{sgSd}. We first notice that
$\gamma^{(K,\xnot)}$ as in  \rf{AetaB} can be generalized
in any dimension, where now $\eta$ is the tautological $(d-3)$-form
on the configuration space of $K\backslash\xnot$.
An explicit computation shows that
\beq
s\gamma^{(K,\xnot)}_\cl = (-1)^{d+1+p}\,
\oint_K [\psi\wedge A + (-1)^{q+d+1}\, B\wedge c],
\quad\mbox{on shell}.
\eeq
Therefore, {\em in odd dimension} we can define the
following on-shell $s$-closed functional
\beq
\gammatotp\cl = \gamma^{\Sigma_K}_\cl + (-1)^{p+1}\,\gamma^{(K,\xnot)}_\cl.
\lab{defgtcld}
\eeq

Then, starting from \rf{defgtcld},
the BV procedure will yield a $\sigma$-closed observable.

Finally, we would like to consider an object like $\calO_0$ in 
\rf{calO0}, for its v.e.v.\ should be related to the Alexander--Conway 
polynomial (or its inverse, depending on $p$). Of course, we do not 
expect $\calO_0$ to be an observable, so we should look
for corrections as explained in subsection \ref{ssec-cspko}.

Notice that in any dimension it is possible to define linear
combinations $\calA$ and $\calB$ generalizing \rf{defcalAB},
\rf{Somegacal} and \rf{sigmacalAB} (and including the whole
set of ghosts for ghosts).

Moreover, in odd dimension the classical action is invariant under 
$(A,B)\to (B,A)$ while $\gammatotp\cl$ is odd under it. Thus, their
BV extensions will share the same property under $(\calA,\calB)\to
(\calB,\calA)$.  This leads to proving a generalization of 
Thm.\ \fullref{th-notan} stating that $\calO_0$ is never anomalous.
We have only to check that the form degree of the one-ghost-number 
component of any form with well-defined parity under the above 
transformation never matches with the dimension of a nontrivial homology
space of $C_n(\reali^d)$. As a matter of fact, these dimensions are
multiples of $(d-1)$. On the other hand, forms with well-defined parity
are obtained by products of $\calB\wedge\calA$ and $\widetilde\calB
\wedge\calA$, both of which are overall $(d-1)$-forms, times 
a certain number $r$ of tautological $(d-3)$-forms $\eta$; thus,
the form degree of the one-ghost-number component will be
conguent to $-1-2r$ mod $(d-1)$. Then our claim follows from the
fact that
\[
2r+1\equiv0\ \rm{mod}(d-1)
\]
has no solutions if $d$ is odd.

The v.e.v.\ of $\calO$ should then yield metric-independent
functionals of the knot and its spanning surface.
Eventually, if a vanishing theorem holds, these functionals will be 
knot invariants.
Hence, we could compute numerical knot
invariants (presumably the coefficients of the Alexander--Conway 
polynomial or its inverse) in any odd dimension in terms of integrals 
over the configuration spaces of points on the knot and on its spanning 
surface.

Via Stokes's theorem,
the second-order invariants should correspond, up to a constant
term, to the ones proposed in \cite{Bott-un}.

As a final remark, we notice that in three dimensions we could have
chosen $A$ to be a zero-form and $B$ a two-form. In this case, the v.e.v.\ of $\calO$ should give directly the Alexander--Conway polynomial 
instead of its inverse. 

\section{Conclusions}
In this paper we have considered a new way of obtaining knot invariants
from a TQFT.

The nice feature of our theory is that it is Abelian.
What makes things non-trivial is a rather involved observable,
which can be defined only in the context of BV formalism;
yet, as observed in the last section, it can be generalized in any odd 
dimension.

In the three-dimensional case, 
we have shown that at the second order the theory actually produces
a numerical knot invariant which, despite the fact it is not new,
comes out written in an entirely new way.

The next task is to find the other corrections to the observable
and to evaluate higher-order v.e.v.'s, in three dimensions as well
as in any odd dimension.

Of course, an alternative way would be working directly on the space of
surface-plus-knot diagrams, as described in subsection \ref{ssec-reg},
and try to find combinations whose differential vanishes. This would
allow studying higher-degree forms on the space of imbeddings as well.

Notice that while the Chern-Simons and the $BF$ theory with
a cosmological term produce the whole set of HOMFLY polynomials
and their ``colored" generalizations, pure $BF$ theories, both Abelian
and non-Abelian, give only the Alexander--Conway polynomial.  

However, it is possible that even more involved observables
exist whose v.e.v.\ is a more general knot invariant. As a matter of
fact, pure $BF$ theory comes out naturally as a particular limit of
the v.e.v.\ of a cabled Wilson loop in the theory with a cosmological
term \cite{Cat}. This limit corresponds to the first diagonal
in the $(h,d)$ expansion of the colored Jones function \cite{MM}. 
A generalization
of the computation done in \cite{Cat} should give the observables
whose v.e.v.'s correspond to the upper diagonals in this expansion.
Then a careful study of the ``Abelianizing limit" described in the
Introduction should yield the corresponding observables for the Abelian
theory. 

%\acknowledgements
\section*{Acknowledgements}
I thank D.~Anselmi, P.~Cotta-Ramusino and R.~Longoni for
helpful conversations. I am especially thankful to R.~Bott 
for a number of very useful discussions.

This work was supported by INFN Grant No.\ 5077/94.

\appendix
\section{The treatment of the harmonic zero modes}
\lab{app-zm}
If the $\omega$-Laplacian 
$\domega^*\domega+\domega\domega^*$ has zero modes,
the formulae in Sec.\ \ref{tdabft} have to be slightly modified.
We will essentially follow the approach explained in \cite{BlT}, 
adapting it to the BV formalism. Notice that we suppose here that
$H^1(M,\domega)$ is not trivial, but we go on assuming that
$H^0(M,\domega)$ is trivial since we want the symmetry group to act freely.

The first step is to modify the BV action \rf{Somega} so as to include
the symmetry obtained by adding an  $\omega$- [($-\omega$)-]harmonic 
form to $A$ [$B$]; viz.,
\beq
S^\omega\longrightarrow S^\omega+\sum_{\alpha=1}^{b_1[\omega]}
\left[{
\int_M(
A^*\wedge k^\alpha\varphi_\alpha + B^*\wedge p^\alpha\varphib_\alpha)
+\bar k^\dg_\alpha \chi^\alpha + \bar p^\dg_\alpha\pi^\alpha
}\right],
\lab{Somega+}
\eeq
where
\begin{itemize}
\item $b_1[\omega]=\dim H^1(M,\domega)$;
\item $\{\varphi_\alpha\}$ [$\{\varphib_\alpha\}$] is an orthonormal
basis of $\omega$- [($-\omega$)-]harmonic one-forms; it is convenient
to choose the normalization
\beq
\braket{\varphi_\alpha}{\varphi_\beta}=
\braket{\varphib_\alpha}{\varphib_\beta}=
v\ \delta_{\alpha\beta},
\eeq
where $v$ is the volume of the manifold $M$ (which we suppose to
be compact);
\item $k^\alpha$ and $p^\alpha$ are constant fields
with ghost number one;
\item $\bar k^\alpha$ and $\bar p^\alpha$ are constant fields
with ghost number minus one, and
\item $\chi^\alpha$ and $\pi^\alpha$ are constant fields
with ghost number zero.
\end{itemize}
Notice that now the action \rf{Somega+} is not metric independent;
as a matter of fact, the choice of the bases 
$\{\varphi_\alpha\}$ and $\{\varphib_\alpha\}$ requires a volume
form. Thus, we cannot expect the partition function to be
a topological invariant. However, it is not difficult to see
that the argument given in Sec.\ \ref{tdabft} to prove that the
partition function depends only on the cohomology class of $\omega$
still holds.

The action of the $\sigma^\omega$ operator is the same as in
\rf{sigmaAB} and \rf{sigmaanti} but on $A$ and $B$ where it acts
as follows:
\beq
\sigmao A = \domega c + \sum_{\alpha=1}^{b_1[\omega]} k^\alpha\varphi_\alpha,
\quad
\sigmao B = \domegab\psi +
\sum_{\alpha=1}^{b_1[\omega]} p^\alpha\varphib_\alpha.
\lab{sigmaAB+}
\eeq
Moreover, we have
\beq
\begin{array}{cc}
\sigmao k^\dg_\alpha = \int_M A^*\wedge\varphi_\alpha, &
\sigmao k_\alpha =0,\\
\sigmao p^\dg_\alpha = \int_M B^*\wedge\varphib_\alpha, &
\sigmao p_\alpha =0,
\end{array}
\eeq
\beq
\begin{array}{cccc}
\sigmao\chi^\dg_\alpha=-\bar k^\dg_\alpha, &
\sigmao\bar k^\dg_\alpha=0, &
\sigmao\bar k^\alpha=\chi^\alpha, &
\sigmao\chi^\alpha=0,\\
\sigmao\pi^\dg_\alpha=-\bar p^\dg_\alpha, &
\sigmao\bar p^\dg_\alpha=0, &
\sigmao\bar p^\alpha=\pi^\alpha, &
\sigmao\pi^\alpha=0.
\end{array}
\eeq

The gauge-fixing fermion defined in \rf{defPsi} has now
to be modified as
\beq
\Psi\longrightarrow\Psi + \sum_{\alpha=1}^{b_1[\omega]}\left({
\bar k^\alpha\braket{\varphi_\alpha}A +
\bar p^\alpha\braket{\varphib_\alpha}B
}\right)
\lab{defPsi+}
\eeq
in order to fix the new symmetries \rf{sigmaAB+}.
With this choice of gauge, we have
\beq
Z[M,\omega] =\int [D\Phi] 
\ \prod_{\alpha=1}^{b_1[\omega]}
dk^\alpha\,d\bar k^\alpha\,dp^\alpha\,d\bar p^\alpha
\,d\chi^\alpha\,d\pi^\alpha
\ \exp{(i S^\omega_{\rm g.f.})},
\lab{Zomega+}
\eeq
where 
\beq
[D\Phi] = [DA\,DB\,Dc\,D\bar c\,D\psi\,D\bar\psi\, 
Dh_c\,Dh_\psi],
\eeq
and $S^\omega_{\rm g.f.}$ is the following modification of
\rf{Somegagf}
\beq
S^\omega_{\rm g.f.}\longrightarrow
S^\omega_{\rm g.f.} +
\sum_{\alpha=1}^{b_1[\omega]}\left[{
v\,(\bar k^\alpha k^\alpha + \bar p^\alpha p^\alpha) +
\chi^\alpha\braket{\varphi_\alpha}A+
\pi^\alpha\braket{\varphib_\alpha}B.
}\right]
\eeq
An explicit computation of \rf{Zomega+}, in zeta-function
regularization, shows that
\beq
Z[M,\omega] = v^{2b_1[\omega]}\,\calT(M,\domega);
\lab{ZomegaT+}
\eeq
thus, apart from a volume factor, the partition function is still
a topological invariant.

\section{The first correction to the observable $\calO_0$}
\lab{app-corr}
In this appendix we solve eqn.\ \rf{Onn} for $n=1$. 
By \rf{DeltaO0}, we have 
\beq
\sigma\calO_1=\frac12\,\calO_0\,
\antib\gammatot\gammatot;
\eeq
thus, by \rf{O1},
\beq
\sigma U_1=\frac12\,\antib\gammatot\gammatot.
\lab{U1}
\eeq
An explicit computation shows that the nonvanishing
terms of \rf{U1} are given by
\beq
2\sigma U_1=
\antib{\intcdue A_1\eta_{12} B_2}
{\intsig [B^*(\psi-\psi(\xnot))+(c-c(\xnot))A^*]}.
\lab{gg}
\eeq
(We suppress all the $\wedge$ symbols for simplicity.)

Since the antibracket contracts fields evaluated at the same point,
we have the following identities
\beq
\begin{array}{lcl}
\antib{\intcdue A_1\eta_{12} B_2}{\intsig B^*(\psi-\psi(\xnot))}
&=&
\antib{\intcdue A_1\eta_{12} B_2(\psi_2-\psi_0)}{\intsig B^*},\\
\antib{\intcdue A_1\eta_{12} B_2}{\intsig (c-c(\xnot))A^*}
&=&
-\antib{\intcdue (c_1-c_0)A_1\eta_{12} B_2}{\intsig A^*};
\end{array}
\lab{id1}
\eeq
moreover, since the antibracket of $A$ with $A^*$ is the same
as the antibracket of $B$ with $B^*$, we have also the following
identities
\beq
\begin{array}{lcl}
\antib{\int A_1\eta_{12} B_2\psi_2}{\int B^*}
&=&
\antib{\int A_1\eta_{12} A_2\psi_2}{\int A^*}-
\antib{\int B_1\eta_{12} A_2\psi_2}{\int B^*},\\
\antib{\int A_1c_1\eta_{12} B_2}{\int A^*}
&=&
\antib{\int B_1c_1\eta_{12} B_2}{\int B^*}-
\antib{\int B_1c_1\eta_{12} A_2}{\int A^*}.
\end{array}
\lab{id2}
\eeq
By \rf{id1} and \rf{id2}, \rf{gg} now reads
\beq
\begin{array}{lcl}
2\sigma U_1 &=&
\antib{\intcdue [A_1\eta_{12}A_2\psi_2 +
B_1c_1\eta_{12}A_2 + c_0A_1\eta_{12}B_2]}
{\intsig A^*} +\\ &-&
\antib{\intcdue [B_1\eta_{12}A_2\psi_2 +
B_1c_1\eta_{12}B_2 + A_1\eta_{12}B_2\psi_0]}
{\intsig B^*}.
\end{array}
\lab{gg2}
\eeq
If now we define $\gamma_{ABB}$ and $\gamma_{BAA}$ as in
\rf{gammaABB} and \rf{gammaBAA}, we see that
\beq
\begin{array}{lcl}
\frac12\,\antib\gammatot\gammatot&=&
-\antib{\sigma\gamma_{ABB}}{\int_{\Sigma_K} B^*}
-\antib{\sigma\gamma_{BAA}}{\int_{\Sigma_K} A^*}=\\
 &=&\Delta\left[{
(\sigma\gamma_{ABB})\,\int_{\Sigma_K} B^* +
(\sigma\gamma_{BAA})\,\int_{\Sigma_K} A^*
}\right].
\end{array}
\eeq
Since
\[
\Delta\left[{
\gamma_{ABB}\,(\sigma\int_{\Sigma_K} B^*) +
\gamma_{BAA}\,(\sigma\int_{\Sigma_K} A^*)
}\right]=0,
\]
we get
\beq
\frac12\,\antib\gammatot\gammatot=
-\Delta\sigma u_1 = \sigma\Delta u_1,
\eeq
where we have used \rf{Deltasigma}, and $u_1$ is defined in
\rf{u1}. By \rf{U1} and \rf{O1}, we arrive at \rf{Uu}
up to a $\sigma$-closed term.   In particular,
the $\Delta$-, $\sigma$-closed correction leading to \rf{aveO2tau}
can be written as
\beq
U_1\rightarrow U_1 -\frac k2\delta U_1 + l\tau+i\lambda m,
\lab{deltaU1}
\eeq
with $\tau$ defined in \rf{deftauK}, and
\beq
\delta U_1=\Delta\delta u_1,
\eeq
where
\beq
\delta u_1 = \gamma_A^K (\gamma_B^K)^2 \intsig B^* +
\gamma_B^K (\gamma_A^K)^2 \intsig A^* .
\eeq
In fact, 
\beq
i\lambda\Delta\delta u_1 = \sigma\delta u_1 - \Omega\delta u_1 =
-2\tau^2 - \Omega\delta u_1;
\eeq
therefore, by Statement \fullref{statBV},
\beq
\ave{i\lambda\Delta\delta u_1}_\lambda =
-2\ave{\tau[K]^2}_\lambda.
\eeq

\section{Vanishing theorems}
\lab{app-hf}
\subsection{The hidden-face contribution to $dV$}
In this subsection we show that $dV$ does not have hidden-face 
contributions. The only nontrivial hidden face is the one where
the three point come together (it is easy to show that the pushforward
of $[V]$ along all other hidden faces vanishes). We will show that 
the push forward of $[V]$ vanishes on this face, too.

To begin with, we parametrize the three points on this face, which
we will denote by $S$, as
\beq
\begin{array}{lcl}
x_1 &=& \Phi(y) + r \frac{\dot\Phi(y)}{|\dot\Phi(y)|} u_1 +\cdots,\\
x_2 &=& \Phi(y) + r \frac{\dot\Phi(y)}{|\dot\Phi(y)|} u_2 +\cdots,\\
x_3 &=& \Phi(y) + r \left({
\frac{\dot\Phi(y)}{|\dot\Phi(y)|} u_3 +
\frac{\Phior(y)}{|\Phior(y)|} u_4}\right)
+\cdots,
\end{array}
\lab{x123}
\eeq
where $y\in S^1$, $r\to 0^+$ and the $u$ variables live in the 
two-dimensional manifold
\beq
\calU=\{
\begin{array}[t]{l}
(u_1,u_2,u_3,u_4)
\in\reali^3\times\reali^+/
u_1<u_2, \sum_{i=1}^4 (u_i)^2 =1,
\sum_{i=1}^3 u_i = 0,\\
(u_i,0)\not=(u_3,u_4), i=1,2
\}.
\end{array}
\eeq
Following the approach of \cite{BT}, we want to represent $[V]$ on
$S$ as the pull back of a universal form $\hat\lambda$. More precisely, 
we introduce the maps
\beq
\psi : 
\begin{array}[t]{ccc}
\Ctilde10 &\rightarrow& S^2\\
y&\mapsto& \frac{\dot\Phi(y)}{|\dot\Phi(y)|}
\end{array},
\eeq
\beq
\chi : 
\begin{array}[t]{ccc}
\Ctilde10 &\rightarrow& S^2\\
y&\mapsto& \frac{\Phior(y)}{|\Phior(y)|}
\end{array},
\eeq
and
\beq
f =(\psi,\chi): \Ctilde10 \rightarrow S^2\times S^2.
\lab{fmap}
\eeq
Then we consider the commutative diagram
\beq
\begin{array}{ccc}
S & \stackrel{\hat f}{\longrightarrow} & \calU\times S^2\times S^2\\
\Big\downarrow\vcenter{\rlap{$\pi$}} & &
\Big\downarrow\vcenter{\rlap{$\hat\pi$}} \\
\Ctilde10 & \stackrel{f}{\longrightarrow} & S^2\times S^2
\end{array},
\lab{commdiag}
\eeq
where $\hat f$ maps $S$ into the parametrization $\calU$ of the blow up
times the value of $f$,
while $\pi$ and $\hat\pi$ are projections along the fibers.
Eventually, we notice that there exists a universal 
four-form $\hat\lambda$ on $\calU\times S^2\times S^2$ such that
\beq
[V] = \hat f^* \hat\lambda\quad\mbox{and}\quad
\pi_*[V] = f^*\hat\pi_*\hat\lambda.
\lab{fstar}
\eeq
To write $\hat\lambda$ explicitly, we have to introduce the maps
$\phi_{13}$, $\phi_{23}$ and $\phi$ defined as
\beq
\phi_{i3} : 
\begin{array}[t]{ccc}
\calU\times S^2\times S^2 &\rightarrow& S^2\\
(u,a,b) &\mapsto& \frac{(u_3-u_i)a - u_4b}{|(u_3-u_i)a - u_4b|}
\end{array},\quad i=1,2,
\lab{defphii}
\eeq
and
\beq
\phi = (\phi_{13},\phi_{23}) : 
\calU\times S^2\times S^2 \rightarrow S^2\times S^2.
\lab{defphi}
\eeq
Then we have
\beq
\hat\lambda = \phi^*(\omega_1\wedge\omega_2),
\eeq
where $\omega_1$ and $\omega_2$ are the $SO(3)$-invariant unit elements
on the two spheres.

Now we notice that $\phi$ and $\hat\pi$ are equivariant maps
under the action of $SO(3)$ (we mean the diagonal action on
$S^2\times S^2$ and the trivial action on $\calU$), and that
$\omega_1\wedge\omega_2$ is $SO(3)$-invariant.
Thus,  $\hat\pi_*\hat\lambda$ is rotationally invariant as well. 
Since it is a two-form and the only 
rotational invariant forms (up to multiplication by a scalar) on $S^2$ 
are $1$ and $\omega$, we conclude that $\hat\pi_*\hat\lambda$ reads
\beq
\hat\pi_*\hat\lambda = c_1 \omega_1 + c_2 \omega_2,
\lab{defhatlambda}
\eeq
where $c_1$ and $c_2$ are constants.

Finally, we consider the (orientation reversing)
automorphism $\Theta$ of $S^2$ that maps a
point into its antipode, and notice that $\Theta^*\omega=-\omega$.
Therefore, if we consider the diagonal extension of $\Theta$ to 
$S^2\times S^2$, we see that
\beq
\Theta^* \hat\pi_*\hat\lambda = -\hat\pi_*\hat\lambda.
\lab{lambdaodd}
\eeq
On the other hand, 
$\phi$ and $\hat\pi$ are equivariant maps under the action of $\Theta$ 
(we mean the diagonal action on $S^2\times S^2$ and the trivial action 
on $\calU$).
Since $\Theta^*(\omega_1\wedge\omega_2)=\omega_1\wedge\omega_2$,
we conclude that
\beq
\Theta^* \hat\pi_*\hat\lambda = \hat\pi_*\hat\lambda.
\lab{lambdaeven}
\eeq
This, together with \rf{lambdaodd}, implies that $\hat\pi_*\hat\lambda$
vanishes. Thus, owing to \rf{fstar}, we have proved the following
\begin{Th}
The push forward of [V] along the hidden face where all the points
collapse together vanishes.
\end{Th}

\subsection{The general vanishing theorem}
A generalization of the previous argument leads to proving a more
general vanishing theorem for diagrams involving tautological forms
connecting points on the knot and/or on its spanning surface. 

The first step is to generalize the commutative diagram \rf{commdiag}.
In general, the stratum $S$ we are considering will have a natural 
projection $\pi$ to a configuration space with less points [which will
replace $\Ctilde 10$ in the lower left corner of \rf{commdiag}].
Suppose we start with $n$ points on the knot and $t$ points on
the surface, i.e., with the $(n+2t)$-dimensional configuration space 
$\Ctilde nt$. We want $q$ points on the knot and $s$ points on the 
surface to collapse. There are three cases:
\begin{description}
\item{Case 1)} $q=0$ and the $s$ points collapse together on the 
surface;  
\item{Case 2)} the points collapse on the base point;
\item{Case 3)} otherwise.
\end{description}
In case 1, $\pi$ projects $S$ to $\Ctilde n{t-s+1}$;
in case 2, to $\Ctilde {n-q}{t-s}$;
in case 3, to $\Ctilde {n-q+1}{t-s}$.
The dimension of the fiber, which we shall denote by $D$, can be 
computed noticing that $\dim S = \dim\Ctilde nt-1$. In case 1, we
have $D=2s-3$; in case 2, $D=q+2s-1$; in case 3, $D=q+2s-2$.

We will prove that the push forward along a face of $\Ctilde nt$
vanishes if $D>0$. Since $D=0$ is satisfied only in case 2, with $(q=1,s=0)$,
or in case 3, with either $(q=0,s=1)$ or $(q=2,s=0)$, we have the 
following
\begin{Th}
The push forward along a hidden face always vanishes; so does the
push forward along a principal face unless it is simple.
\lab{th-vt}
\end{Th}
(For a definition of principal, hidden and simple principal faces, s.\ 
page~\pageref{simple}.)

In particular, this proves that the contributions $\ell_l$, $\ell_r$
and $\ell_0$---considered in subsection \ref{ssec-isot}---vanish.

\paragraph{Proof}
First we split the form on $S$ into the product 
$\lambda_1\wedge\pi^*\lambda_2$. 
The propagators that define $\lambda_2$ connect either two points that
do not collapse, or a point that does not collapse with a point that 
does, or two of the $q$ points on the knot that collapse together.
The propagators that define $\lambda_1$ connect two collapsing points
at least one of which is on the surface.

It follows that $\lambda_1$ is written in terms of products
of pull backs of $\omega$ via maps that are linear combinations of
the unit vectors $\dot\Phi/|\dot\Phi|$ and $\Phior/|\Phior|$.
Thus, $S$ will be mapped to a proper 
submanifold of $(S^2)^n$ unless $n=0$, $n=1$ or $n=2$. This means that 
$\lambda_1$ vanishes unless it is a zero-, a two- or a four-form. 
In the first case, however,  $\pi_*\lambda_1$ clearly vanishes unless $D=0$.

Now the idea is to generalize the commutative diagram \rf{commdiag}
and write $\lambda_1$ in terms of a universal form
$\hat\lambda_1$:
\beq
\lambda_1 = \hat f^* \hat\lambda_1\quad\mbox{and}\quad
\pi_*\lambda_1 = f^*\hat\pi_*\hat\lambda_1.
\lab{fstar1}
\eeq
The left column of \rf{commdiag} is unchanged if we still denote
by $\calU$ the parametrization of the blow up. Explicitly
the manifold $\calU$ reads
\beq
\calU=\{
\begin{array}[t]{l}
((u_1^1,u_1^2),\ldots,(u_s^1,u_s^2))\in(\reali\times\reali^+)^s
\ / \ \sum_{i=1}^s\sum_{\alpha=1}^2(u_i^\alpha)^2=1,\ 
\sum_{i=1}^s u_i^\alpha=0,\\
u_i^\alpha\not=u_j^\alpha\ \forall\alpha \mbox{
if $i$ and $j$ are connected}
\},
\end{array}
\lab{ucase1}
\eeq
in case 1;
\beq
\calU=\{
\begin{array}[t]{l}
(u_1,\ldots,u_q,
(u_{q+1}^1,u_{q+1}^2),\ldots,(u_{q+s}^1,u_{q+s}^2))
\in\reali^q\times(\reali\times\reali^+)^s\ /\\
u_1<\cdots<u_q,\ 
\sum_{i=1}^q (u_i)^2 +
\sum_{i=q+1}^{q+s}\sum_{\alpha=1}^2(u_i^\alpha)^2=1,\\
u_i^\alpha\not=u_j^\alpha\ \forall\alpha \mbox{ and }
(u_i,0)\not=(u_j^1,u_j^2) \mbox{
if $i$ and $j$ are connected}
\},
\end{array}
\lab{ucase2}
\eeq
in case 2, and
\beq
\calU=\{
\begin{array}[t]{l}
(u_1,\ldots,u_q,
(u_{q+1}^1,u_{q+1}^2),\ldots,(u_{q+s}^1,u_{q+s}^2))
\in\reali^q\times(\reali\times\reali^+)^s\ /\\
u_1<\cdots<u_q,\ 
\sum_{i=1}^q (u_i)^2 +
\sum_{i=q+1}^{q+s}\sum_{\alpha=1}^2(u_i^\alpha)^2=1,\\
\sum_{i=1}^q u_i + \sum_{i=q+1}^{q+s} u_i^1=0,\\
u_i^\alpha\not=u_j^\alpha\ \forall\alpha \mbox{ and }
(u_i,0)\not=(u_j^1,u_j^2) \mbox{
if $i$ and $j$ are connected}
\},
\end{array}
\lab{ucase3}
\eeq
in case 3.

The universal form $\hat\lambda_1$ is written in terms of
pullbacks of $\omega$ via $\hat\phi$-maps defined as follows:
\beq
\hat\phi_{ij} = \univec{(u_j^1-u_i)a+u_j^2b},
\quad\mbox{if $i\leq q$ and $q<j\leq q+s$},
\lab{phiijks}
\eeq
\beq
\hat\phi_{ij} = \univec{(u_j^1-u_i^1)a+(u_j^2-u_i^2)b},
\quad\mbox{if $q<i<j\leq q+s$},
\lab{phiijss}
\eeq
and $\hat\phi_{ji}=-\hat\phi_{ij}$.

We must now distinguish two subcases: viz., when
$\hat\lambda_1$ is a two-form and when it is a four-form.

In the first subcase, $\hat\lambda_1$ is obtained via pull back
through a single $\hat\phi$-map to $S^2$.  
Rotational invariance shows that 
$\hat\pi_*\hat\lambda_1$ does not vanish only if it is a zero- or a
two-form; in the former instance it is a constant, in the latter
it is linear combination of $\omega_1$ and $\omega_2$. 
However,
$\hat\pi_*\hat\lambda_1$ must be odd under the action of the
automorphism $\Theta$ that maps a point on $S^2$ into its antipode
since $\omega$ is odd.
Therefore, it does not vanish only in the
latter instance. However, if both $\hat\lambda_1$ and 
$\hat\pi_*\hat\lambda_1$ are two-forms, we have $D=0$.

In the second subcase, $\hat\lambda_1$ is obtained via pull back
through a product of two $\hat\phi$-maps to $S^2\times S^2$, 
as, e.g., in \rf{defphi}. Rotational invariance shows that
$\hat\pi_*\hat\lambda_1$ does not vanish only if it is a zero-, a
two- or a four-form; in the first instance it is a constant, in the 
second it is a linear combination of $\omega_1$ and $\omega_2$, in the 
third it is a multiple of $\omega_1\wedge\omega_2$. 
However,
$\hat\pi_*\hat\lambda_1$ must be even under the diagonal action of the
automorphism $\Theta$ since $\omega_1\wedge\omega_2$ is even.
Thus, $\hat\pi_*\hat\lambda_1$  
does not vanish only if it is a zero- or a four-form. Since 
$\hat\lambda_1$ is a four form, we have $D=4$ in the former instance and 
$D=0$ in the latter.

Therefore, to complete the proof we have only to show that, in the 
former instance, $\hat\pi_*\hat\lambda_1$ vanishes. This happens since,
if $D=4$, $\hat\pi_*$ selects the $(4,0)$-component of the four-form
$\hat\lambda_1$ on $\calU\times(S^2\times S^2)$. This component, 
however, vanishes since, for fixed $(a,b)\in S^2\times S^2$, $\hat\phi$ 
maps the four-dimensional manifold $\calU$ into a two-dimensional 
submanifold of $S^2\times S^2$ [this submanifold is parametrized by two 
unit vectors $z_i(u;a,b)$ satisfying $z_i\cdot (a\times b)=0,\ i=1,2$].

This concludes our proof.

%\begin{references}
\end{document}